\documentclass[aps,showpacs,amsmath,amssymb,prb,groupedaddress,preprint]{revtex4-1}
\usepackage{color}
\usepackage{graphicx,dcolumn,diagbox,bm}
\newcommand{\MCO} {{\mbox{MgCr${}_2$O${}_4$}}}
\newcommand{\ZCO} {{\mbox{ZnCr${}_2$O${}_4$}}}
\newcommand{\ACO} {{\mbox{ACr${}_2$O${}_4$}}}

\newcommand{\DSC} {({\ensuremath{\frac{d\sigma}} {d\Omega}})}
\newcommand{\half}{{\ensuremath{\frac{1}{2}}}}
\newcommand{\third}{{\ensuremath{\frac{1}{3}}}}
\newcommand{\threehalf}{{\ensuremath{\frac{3}{2}}}}
\newcommand{\fivehalf}{{\ensuremath{\frac{5}{2}}}}

\newcommand{\Ps}{\mbox{\bf P}} 
\newcommand{\Pin}{\mbox{\bf P$'$}} 
\definecolor{green}{rgb}{0.15,0.85,0.35}

\begin{document}

\title{Manifolds of magnetic ordered states and excitations in the almost Heisenberg pyrochlore antiferromagnet MgCr$_2$O$_4$}

\author{S. Gao}
\affiliation{Laboratory for Neutron Scattering and Imaging, Paul Scherrer Institute, CH-5232 Villigen, Switzerland}
\author{K. Guratinder}
\affiliation{Laboratory for Neutron Scattering and Imaging, Paul Scherrer Institute, CH-5232 Villigen, Switzerland}
\author{V. Tsurkan}
\affiliation{Experimental Physics V, Center for Electronic Correlations and Magnetism, University of Augsburg, D-86159 Augsburg, Germany}
\affiliation{Institute of Applied Physics, Academy of Sciences of Moldova, MD-2028 Chisinau, Republic of Moldova}
\author{A. Loidl}
\affiliation{Experimental Physics V, Center for Electronic Correlations and Magnetism, University of Augsburg, D-86159 Augsburg, Germany}
\author{M. Ciomaga Hatnean}
\affiliation{Department of Physics, University of Warwick, Coventry CV4 7AL, UK}
\author{G. Balakrishnan}
\affiliation{Department of Physics, University of Warwick, Coventry CV4 7AL, UK}
\author{S. Raymond}
\affiliation{Univ. Grenoble Alpes, CEA, INAC, MEM, F-3800 Grenoble, France}
\author{L. Chapon}
\affiliation{Institut Laue-Langevin, 156X, 38042 Grenoble C\'{e}dex, France}
\altaffiliation[Presently at ]{Diamond Light Source, Didcot OX11 0DE, UK}
\author{V. O. Garlea}
\affiliation{Neutron Scattering Division, Oak Ridge National Laboratory, Oak Ridge, Tennessee 37831, USA}
\author{A. T. Savici}
\affiliation{Neutron Scattering  Division, Oak Ridge National Laboratory, Oak Ridge, Tennessee 37831, USA}
\author{U. Stuhr}
\affiliation{Laboratory for Neutron Scattering and Imaging, Paul Scherrer Institute, CH-5232 Villigen, Switzerland}
\author{J. S. White}
\affiliation{Laboratory for Neutron Scattering and Imaging, Paul Scherrer Institute, CH-5232 Villigen, Switzerland}
\author{M. Mansson}
\affiliation{Laboratory for Neutron Scattering and Imaging, Paul Scherrer Institute, CH-5232 Villigen, Switzerland}
\altaffiliation[Presently at ]{KTH Royal Institute of Technology, Materials Physics, Department of Applied Physics, Stockholm, Sweden}
\author{B. Roessli}
\affiliation{Laboratory for Neutron Scattering and Imaging, Paul Scherrer Institute, CH-5232 Villigen, Switzerland}
\author{A. Cervellino}
\affiliation{Swiss Light Source, Paul Scherrer Institute, CH-5232 Villigen, Switzerland}
\author{A. Bombardi}
\affiliation{Diamond Light Source, Didcot, OX11 0DE, UK}
\author{D. Chernyshov}
\affiliation{Swiss-Norwegian Beamlines at ESRF, F-38000 Grenoble, France}
\author{T. Fennell}
\affiliation{Laboratory for Neutron Scattering and Imaging, Paul Scherrer Institute, CH-5232 Villigen, Switzerland}
\author{Ch. R\"{u}egg}
\affiliation{Laboratory for Neutron Scattering and Imaging, Paul Scherrer Institute, CH-5232 Villigen, Switzerland}
\affiliation{Department of Quantum Matter Physics, University of Geneva, CH-1211 Geneva, Switzerland}
\altaffiliation[Presently at ]{Research Division Neutrons and Muons, Paul Scherrer Institute, CH-5232 Villigen, Switzerland}
\author{J. T. Haraldsen}
\affiliation{Department of Physics, University of North Florida, Jacksonville, Florida 32224, USA} 
\author{O. Zaharko}
\affiliation{Laboratory for Neutron Scattering and Imaging, Paul Scherrer Institute, CH-5232 Villigen, Switzerland }

\date{\today}

\begin{abstract}
In spinels {\ACO} (A=Mg, Zn) realisation of the classical pyrochlore Heisenberg antiferromagnet model is complicated by a strong spin-lattice coupling: the  extensive degeneracy of the ground state is lifted by a magneto-structural transition at $T_N$=12.5 K. We study the resulting low-temperature low-symmetry crystal structure by synchrotron x-ray diffraction. The consistent features of x-ray low-temperature patterns are explained by the tetragonal model of Ehrenberg $et.~al$ [Pow. Diff. $\bf{17}$, 230( 2002)], while other features depend on sample or cooling protocol. Complex partially ordered magnetic state is studied by neutron diffraction and spherical neutron polarimetry. Multiple magnetic domains of configuration arms of the propagation vectors $\bf{k_1}$=({\half} {\half} 0), $\bf{k_2}$=(1 0 {\half}) appear. The ordered moment reaches 1.94(3) $\mu_B$/Cr$^{3+}$ for $\bf{k_1}$
and  2.08(3)$\mu_B$/Cr$^{3+}$ for $\bf{k_2}$, if equal amount of the $\bf{k_1}$ and $\bf{k_2}$ phases is assumed. 
The magnetic arrangements have the dominant components along the [110] and [1-10] diagonals and a smaller $c$-component. By inelastic neutron scattering we investigate the spin excitations, which comprise a mixture of dispersive spin waves propagating from the magnetic Bragg peaks and resonance modes centered at equal energy steps of 4.5 meV. 
We interpret these as acoustic and optical spin wave branches, but show that the neutron scattering cross sections of transitions within a unit of two corner-sharing tetrahedra match the observed intensity distribution of the resonances.
The distinctive fingerprint of cluster-like excitations in the optical spin wave branches suggests that propagating excitations are localized by the complex crystal structure and magnetic orders.\\
\end{abstract}

\maketitle
\section{Introduction}{\label{Sec1}}
One of the long-standing puzzles in frustrated magnetism is the family of {\ACO} spinel chromites (A=Mg, Zn), 
where antiferromagnetically coupled Cr$^{3+}$ ions residing at the vertices of  corner sharing tetrahedra form a highly frustrated pyrochlore lattice. 
A remarkable feature of the excitation spectrum of the chromites is the existence of flat bands dominating the low-energy
excitation spectrum, so-called resonance modes.
The first observation of these resonances is dated to 2002, when Lee $et~al.$ \cite{Lee2002} measured the neutron scattering form factor of the lowest-energy 
excitation in {\ZCO} and recognised that its Fourier transform corresponds to an antiferromagnetic hexagonal spin loop.
The measurement was performed just above the ordering temperature, in the cooperative paramagnet regime, and the hexagonal spin loops were interpreted as local zero energy modes of the pyrochlore Heisenberg antiferromagnet. Such modes  correspond to local correlations of the system as it fluctuates within the ground state manifold.\cite{Chalker1998} 
\\
While studying the excitation spectrum of {\MCO} below $T_N$=12.5 K, Tomiyasu $et~al.$\cite{Tomiyasu2008, Tomiyasu2013} observed a quasi-dispersionless mode at 4.5 meV with the same hexagon-loop form factor, and three further flat bands equally spaced by  $\Delta$E=4.5 meV, that they named resonances. They noticed that the Fourier transform of the higher-energy excitations corresponded to a heptamer, a cluster of 'two corner-sharing tetrahedra' (we abbreviate it to TCST) and suggested that these modes emerge due to  high degeneracy of the excited states. The key questions - do these resonances arise from the zero modes and why are they located at equal energy intervals - remained unsolved.\\
Usually such questions can be answered when the leading terms of the Hamiltonian are identified. This has not yet been achieved for the spin-lattice coupled {\ACO} spinels due to contradictory information about the ground state and poor knowledge of the complete low-energy excitation spectrum.
We therefore first performed detailed synchrotron x-ray and neutron diffraction studies (including powder diffraction, single crystal diffraction with and without magnetic field, spherical neutron polarimetry, Section \ref{Sec2Sub3}- \ref{Sec2Sub5}) to acquire information about the low-temperature crystal structure and long-range magnetic arrangements of {\MCO}, taking it as a representative of the chromite family.  
Secondly, with inelastic neutron scattering we comprehensively measured the low-energy excitation spectrum of {\MCO} single crystals (Section \ref{Sec2Sub6}). We clarify common and individual features of long-wavelength spin waves and resonance modes by performing XYZ-polarization analysis, and measuring temperature and magnetic field dependences.
We derive analytically the inelastic neutron cross sections of the excitations of a TCST cluster by decomposing it into smaller units (Section \ref{Sec3Sub2}). The match between the calculated and observed intensity distributions grants a new view on the origin of the resonance modes.

\section{Experimental facts}{\label{Sec2}}
\subsection{Previous experimental reports}{\label{Sec2Sub1}}
We briefly outline the experimental findings on the {\ACO} chromites (A=Mg, Zn), referring only to the small part of the vast literature on the subject that is relevant to our study.
Magnesium and zinc chromium oxides show very similar magnetic properties with Curie-Weiss temperatures $\Theta_{CW}\approx$ -- 400 K and spin-Peierls transitions at $T_N$ = 12.5 K. This transition is strongly first order. The
magnetic frustration of the pyrochlore lattice is released by distorting regular Cr$^{3+}$ tetrahedra in the high-temperature (HT) phase and thus making the magnetic interactions between Cr$^{3+}$ ions inequivalent. The associated atomic displacements in the low-temperature (LT) phase are small, but sufficient to introduce couplings of sufficiently different strengths.\\
For the LT crystal structure several models have been proposed. 
For {\ZCO} the most detailed LT model is the one published by Ji $et~al.$\cite{Lee2009} It is based on synchrotron x-ray single crystal diffraction data, which contain 140 weak superstructure reflections of the propagation vector $\bf{k}$=({\half} {\half} {\half}).\cite{note1} 
This LT model comprises three types of tetrahedra:
two symmetrically distorted - with even number of short and long bonds, and the third one asymmetrically distorted - with one 
strong and five weak bonds. Such rearrangement of the atoms reduces the magnetic frustration, but only partially - the match between the distortions and the moment arrangement in tetrahedra is incomplete.
For {\MCO} the LT structure has tetragonal ($I4_1/amd$) or orthorhombic ($Fddd$) symmetry\cite{Ehrenberg2002, Ortega2008, Kemei2013} with compression along $c$ and expansion in the $ab$ plane. Kemei $et~al.$\cite{Kemei2013} did not detect the superstructure reflections in synchrotron x-ray powder diffraction patterns, but observed  splitting of the HT cubic reflections, which was interpreted as coexistence of two phases - with tetragonal and orthorhombic symmetries. We suspect that such diversity of observations is caused by sensitivity of the LT structure to the microstructure of the samples (nonstoichiometry, site disorder, defects, etc.), in accord with the  magnetic properties, which are very sensitive to nonstoichiometry.\cite{Dutton2011}\\
Neutron powder and single crystal diffraction reports on magnetic ordering of {\ZCO} and {\MCO} are also controversial.\cite{Plumier1969, Shaked1970, Lee2009} To index magnetic reflections arising at $T_N$ several propagation vectors of the cubic spinel unit cell are necessary.
For {\MCO} Plumier and Sougi\cite{Plumier1969} observed $\bf{k_1}$=({\half} {\half} 0) magnetic reflections and suggested a coplanar structure with magnetic moments in the $ab$ plane. Shaked $et~al.$\cite{Shaked1970} reported two consequent phase transitions with $T_{N1}$=16 K and $T_{N2}$=13.5 K in powder patterns. The magnetic reflections appearing at $T_{N1}$ were indexed with the $\bf{k_5}$=0 wave vector and the ones appearing at $T_{N2}$ with $\bf{k_4}$=({\half} {\half} {\half}). In this study magnetic intensities varied for powder samples, while in a single crystal only the $\bf{k_4}$-phase was present. The authors tested solutions with the most symmetric Shubnikov magnetic space groups and magnetic moments along the three principal directions: $<$100$>$, $<$110$>$, $<$111$>$. The two best models are built from chains of magnetic moments, in the first case the chains propagate along the $ac$- and $ab$- axes, while for the second model - along the $ac$- and $bc$- axes. However, these models could not be distinguished from the available data. An ordered magnetic moment of 2 $\mu_{B}$/Cr$^{3+}$ was obtained, which is less then 3 $\mu_B$, the expected moment of the Cr$^{3+}$ ions.\\
For {\ZCO} two main magnetic wave vectors $\bf{k_1}$=({\half} {\half} 0) and $\bf{k_2}$=(1 0 {\half}) are reported.\cite{Lee2009} Reflections of the $\bf{k_3}$=(100) and $\bf{k_4}$=({\half} {\half} {\half}) wave vectors are weaker
and their intensities vary from sample to sample. Ji $et~al.$\cite{Lee2009} proposed a model for the magnetic structure based on neutron powder diffraction data. Several constraints limiting the number of solutions were employed. Two of them - the same moment magnitude for all Cr$^{3+}$ ions and the zero net magnetic moment for each tetrahedron - are the pillars of the pyrochlore Heisenberg antiferromagnet model. 
The next restriction - confinement of the ordered moment to the tetragonal basal $ab$-plane - results from a polarized neutron experiment.\cite{Lee2008} 
Finally, it was postulated that the magnetic arrangement is a superposition of the $\bf{k_1}$ and $\bf{k_2}$ collinear components. 
The resulting magnetic structure is coplanar with the ordered moment 2.3(2) $\mu_B$/Cr$^{3+}$.
This model reconciles the distortions of  tetrahedra and spin arrangements, though the correspondence is only partial.\\
We also briefly mention the published results on the magnetic excitations in chromites. Inelastic neutron scattering (INS) was measured on {\ZCO} powders\cite{Lee2000} and {\ACO} (A=Zn, Mg) single crystals\cite{Lee2002, Tomiyasu2008, Tomiyasu2013} with the main focus on the resonances. The existence of dispersive spin waves has been mentioned\cite{Lee2000}, but not studied in detail.\\
\subsection{Sample preparation}{\label{Sec2Sub2}}
Our {\MCO} polycrystalline material was synthesized by a solid state reaction between stoichiometric amounts of MgO and Cr$_2$O$_3$ in air. Single crystals were grown by two different methods:
$i$) chemical transport ($ct$) and $ii$) floating zone ($fz$). Perfect single crystals of octahedral shape with dimensions up to 5 mm on the edge were obtained by the first method. They have the cubic spinel structure at room temperature and no inversion between the Mg$^{2+}$ and Cr$^{3+}$ ions according to x-ray diffraction analysis. The boules grown by the floating zone method are up to 70 mm long (along the $a$-axis) and have a diameter of 4-5 mm. Characterisation of the $fz$-samples was more problematic. High-resolution synchrotron x-ray powder diffraction patterns from crushed crystals have cubic symmetry but  the intensity distribution varied from sample to sample and did not match the normal spinel structure and we could not study the degree of inversion. We tolerated these sample peculiarities in order to use large crystals for inelastic neutron scattering, but checked that the observations which are crucial for our conclusions (i. e. magnetic wave vectors, spin waves and resonance modes) also occur in the $ct$-crystals.\\
We used smaller $ct$-crystals to study the low-temperature crystal and magnetic structures, as they have sizes appropriate for diffraction experiments. In addition, six $ct$-crystals were co-aligned into a multi-crystal sample of 250 mg and used for INS experiments on the TASP and EIGER spectrometers at SINQ. To see the fine details of magnetic excitations large mass samples were needed.
Therefore $fz$-crystal boules were co-aligned in a 20-30 mm long 2 g sample and used for INS experiments on HYSPEC at SNS and on IN12 at ILL.
\subsection{Crystal structure below $T_N$}
\begin{figure}
\centering 
\includegraphics[width=0.75\columnwidth]{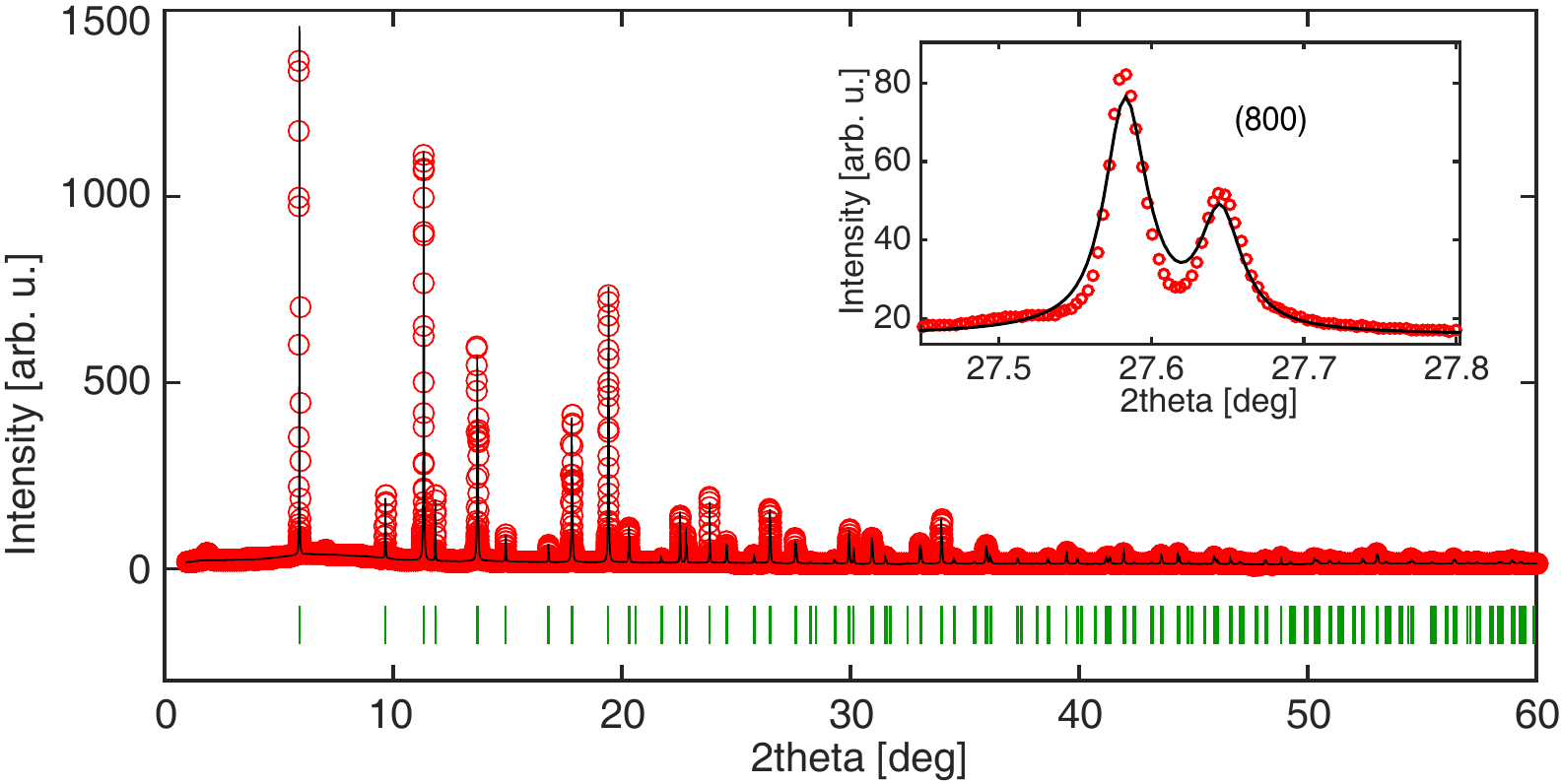} 
\caption{Refinement for the 5.5 K synchrotron diffraction data of {\MCO} with the $I4_1/amd$ space group. Data points are red circles, the calculated pattern is the black solid line, the vertical bars mark the Bragg peaks. The blue line at the bottom is the difference of the data and calculated intensities. The inset zooms the region around the cubic (800) reflection.}
\label{fig:ms_lt} 
\end{figure}

{\label{Sec2Sub3}}
We performed a number of powder and single crystal experiments on several synchrotron x-ray diffraction beamlines to investigate the LT crystal structure of {\MCO}.
Fig. \ref{fig:ms_lt} presents a powder diffraction pattern from a piece of crushed $ct$-crystal collected on the MS beamline of SLS at T=5.5 K.
As can be seen in the inset of Fig. \ref{fig:ms_lt}, the  (800) reflection in the cubic phase splits into two peaks, which indicates a tetragonal distortion. 
The refinement is performed\cite{rodriguez1993} in the $I4_1/amd$ space group\cite{Ehrenberg2002}, and the refined parameters are listed in Table \ref{tab:ms_ref}. 
The results for the $Fddd$ space group with a larger number of refined parameters are equally good.\\
\begin{table}
\caption[Refinement results for the MgCr$_2$O$_4$ structure]{Refinement results for the {\MCO} powder synchrotron x-ray diffraction. The data above the spin-Peierls transition are measured at $T$ = 20 K, below the transition - at 5.5 K.}
\label{tab:ms_ref}
\centering
\small
\begin{tabular}{cccccccccc}
\toprule
 & $a$, \AA & $b$, \AA & $c$, \AA & Mg & Cr & O & $R_p$ & $R_{wp}$ & $\chi^2$\\
 \hline
$Fd\overline{3}m$	& 8.3196 & $-$ & $-$ & $8a$ & $16d$ & $32e$ (0.261 0.261 0.261) & 15.1 & 18.8 & 27.09\\
$I4_1/amd$	& 5.8852 & $-$ & 8.3045 & $4b$ & $8d$ & $16h$ (0.000 0.521 0.739) & 12.3 & 12.8 & 19.86\\
$Fddd$	& 8.3242 & 8.3237 & 8.3059 & $8a$ & $16d$ & $32h$ (0.259 0.264 0.260) & 14.3 & 12.9 & 14.55\\
\toprule
\end{tabular}
\end{table}
In order to check whether superstructure reflections appear in {\MCO} below $T_N$, single crystal synchrotron diffraction data were collected on the Swiss-Norwegian bending-magnet beamlines at ESRF. Only four ${\bf{k}}$=({\half} {\half} {\half}) type reflections were observed and all of them had intensities less than $0.5\times10^{-3}$ of the strongest (111) reflection. In another experiment on the I16 undulator beamline at the Diamond synchrotron, on a larger crystal of the same $ct$-batch, up to 360 superstructure reflections could be measured below $T_N$. Their intensity was also at least $10^{-3}$ times weaker than the intensity of main nuclear reflections. However, refinement of the model of Ji $et~al.$\cite{Lee2009} was not successful. Furthermore, during this experiment several reflections breaking the $F$-centering ($P$-reflections) were detected. They existed above $T_N$ and gained intensity when the crystal was cooled to lower temperatures.\\
Due to the discrepancy between the powder data and single crystal data collected on different crystals and different beamlines we argue that the occurrence of the ${\bf{k}}$=({\half} {\half} {\half}) and $P$-reflections depends on sample or cooling protocol. The tetragonal model of Ehrenberg $et~al.$\cite{Ehrenberg2002} is the highest symmetry and simplest model, which explains consistent features of our all diffraction data. We therefore consider it presently as the best model for the LT structure, though additional static distortions are obviously present.  It is desired to understand better the influence of the microstructure on the LT crystal structure and on the magnetic order discussed below.\\
\subsection{Magnetic ordering}
\begin{figure}
\centering 
\includegraphics[width=0.75\columnwidth]{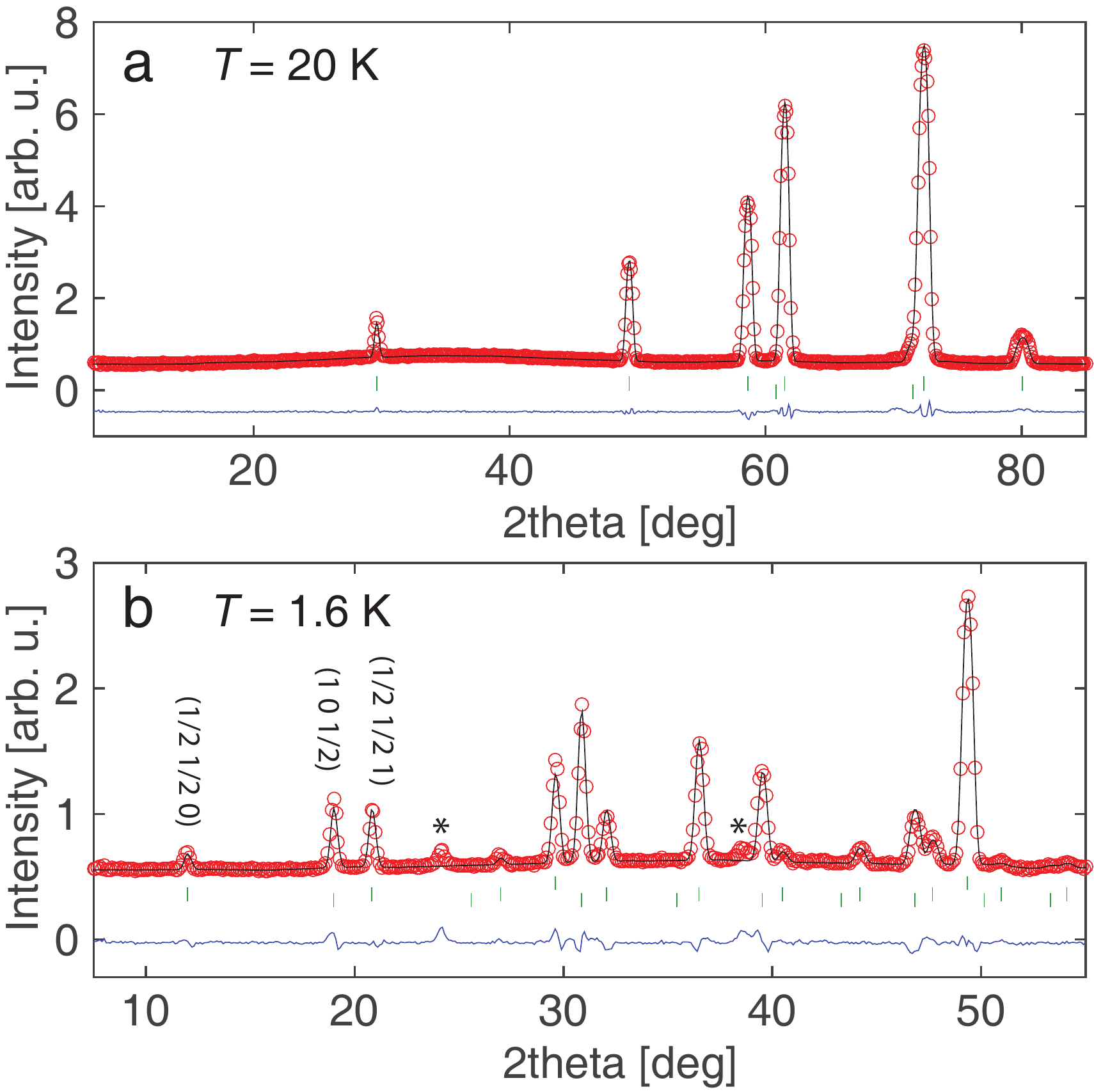} 
\caption{Powder neutron diffraction measured on DMC. Data points are red circles, the calculated pattern is the black solid line, the vertical green bars show the positions of the Bragg peaks. a) 20 K, bars mark Al (can) and nuclear peaks, b) 1.6 K, bars mark nuclear, $\bf{k_1}$=({\half} {\half} 0) and $\bf{k_2}$=(1 0 {\half}) reflections; stars indicate two $\bf{k_3}$=(1 0 0) peaks. The blue line at the bottom is the difference of the data and calculated intensities.}
\label{fig:dmc} 
\end{figure}

\begin{figure}
\includegraphics[width=0.49\columnwidth,keepaspectratio=true]{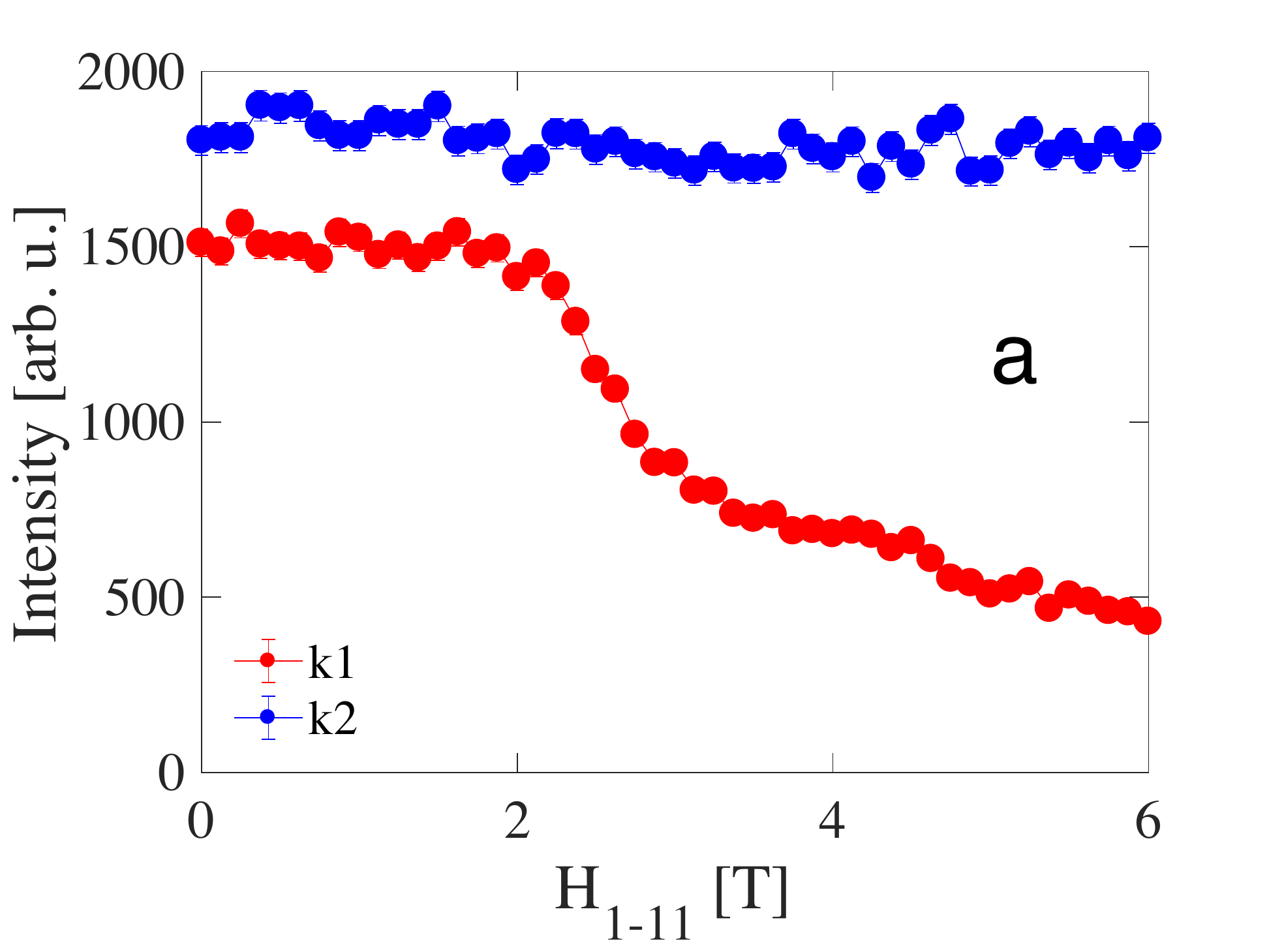}
\includegraphics[width=0.49\columnwidth,keepaspectratio=true]{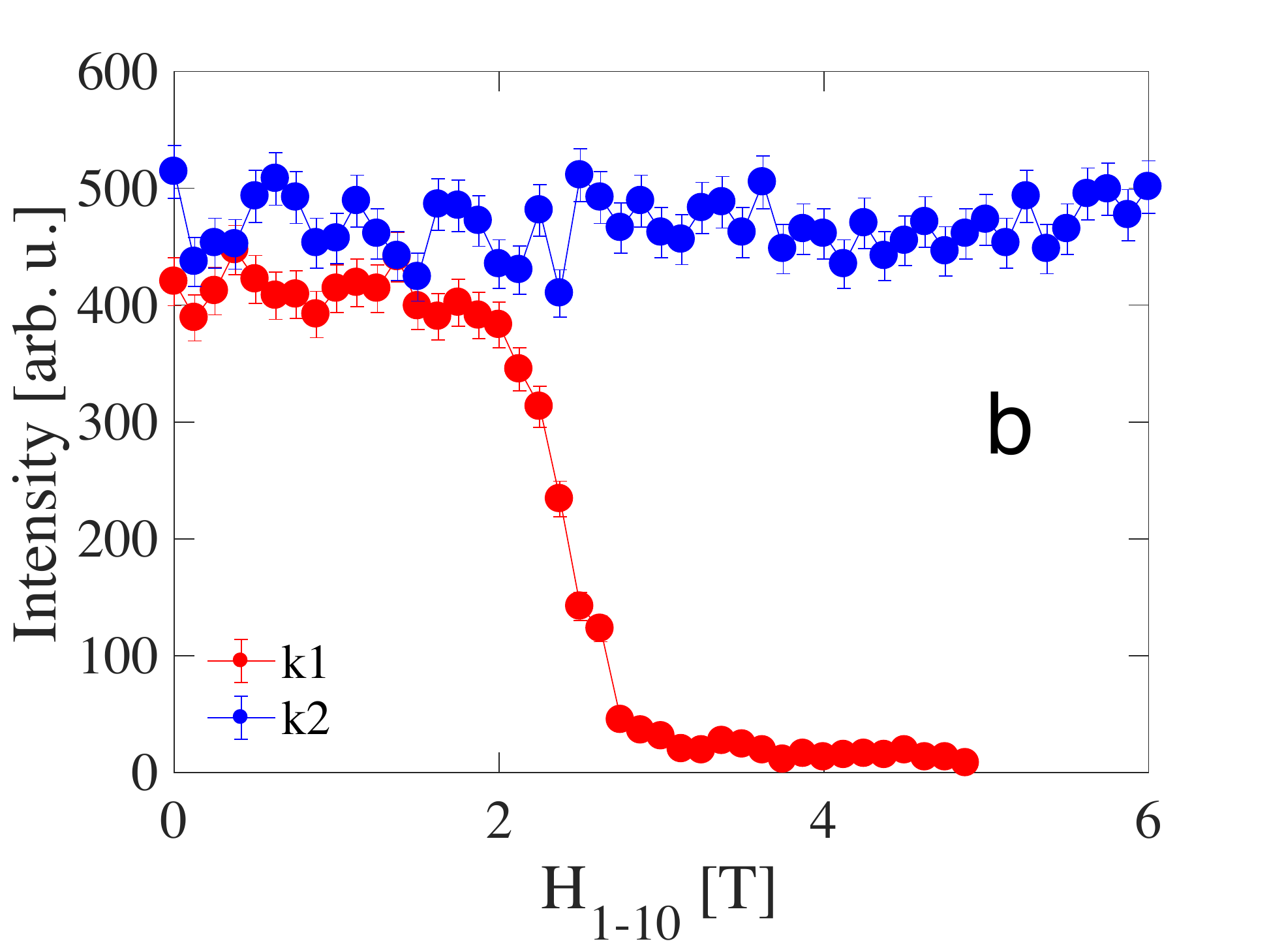}
\caption{Magnetic field dependence of the selected $\bf{k_1}$ and $\bf{k_2}$ reflections at 1.8 K for the {\MCO} crystal cooled in zero magnetic field. a) ({\threehalf} 0 -{\threehalf}) and ({\threehalf} 0 -1) with field applied along ${\bf{H}}_{1-11}$. b) ({\threehalf} {\threehalf} 0) and ({\threehalf} 1 0) with ${\bf{H}}_{1-10}$.}
\label{fig_mf}
\end{figure}

{\label{Sec2Sub4}}
We used a set of neutron diffraction techniques - powder diffraction, single crystal diffraction with and without magnetic field, spherical  neutron polarimetry - to study the magnetic structure of the ground state of {\MCO}.\\
Powder diffraction was measured with the neutron wavelength of 2.5 \AA~on the DMC diffractometer at SINQ. Significant diffuse scattering was detected above the ordering temperature $T_N$ (Fig. \ref{fig:dmc}a). The two main magnetic wave vectors $\bf{k_1}$=({\half} {\half} 0) and $\bf{k_2}$=(1 0 {\half}) and few weak peaks of $\bf{k_3}$=(1 0 0)\cite{note3} were observed below $T_N$ (Fig. \ref{fig:dmc}b). Powder patterns give fast, large-angle coverage of reciprocal space, but overlap of different reflections with the same $2\theta$ due to powder averaging limits the obtained information.\\
Single crystal diffraction data were collected on the TriCS diffractometer at SINQ ($\lambda$=1.178 \AA, 167 reflections of $\bf{k_1}$ and 114 reflections of $\bf{k_2}$) and the D9 diffractometer at ILL ($\lambda$=0.842 \AA, 374 reflections of $\bf{k_1}$ and 270 reflections of $\bf{k_2}$). In these data reflections with the same $2\theta$ are disentangled, but other complications become important - intensities of different twins and different magnetic domains might contribute to the same magnetic reflection.
In the present case with a cubic HT phase three structural twins for the tetragonal LT phase and twelve arms for each of the $\bf{k_1}$ and $\bf{k_2}$ magnetic wave vectors are possible.  
For a nontwinned crystal the configuration\cite{Brown1993} arms with cycling components (i.e. ({\half} {\half} 0), ({\half} 0 {\half})) give rise to separate sets of reflections. For a twinned case they will overlap. The orientation arms with permuting signs (i. e. ({\half} {\half} 0), ({\half} -{\half} 0)) contribute to the same reflections.
The arms of $\bf{k_1}$ and $\bf{k_2}$ might give rise to $n_1$, $n_2$ separate domains (0$<n_i<$12, $i$=1,2); they might combine into a multi-$n_1\bf{k_1}$-$n_2 \bf{k_2}$ structure, or an intermediate case might occur.
It is impossible to distinguish these cases from integrated intensities collected from a single crystal. We used  neutron diffraction in magnetic field and neutron spherical polarimetry to establish that in {\MCO} crystals a multi-domain state appears. Determination of the $\bf{k_1}$ and  $\bf{k_2}$ magnetic structures is presented in Section \ref{Sec2Sub5}.\\
A single crystal diffraction experiment with applied magnetic field $\bf{H}$ was performed on the ZebRa diffractometer at SINQ ($\lambda$=2.317 \AA). Figures \ref{fig_mf} a, b present the behaviour of several reflections with $\bf{H}$ along [1-11] and [1-10], respectively. Intensities of the measured $\bf{k_1}$ reflections change between 2.5 - 3 T, while for $\bf{k_2}$ no changes are detected.\cite{note2} 
We conclude thus, that $\bf{k_1}$ and $\bf{k_2}$ do not build a common multi-$\bf{k}$ structure, but form different domains.\\
Small sets of magnetic reflections of several arms of $\bf{k_1}$ and $\bf{k_2}$ accessible in the normal beam geometry were measured and selected observations are
presented in Table \ref{tab_dom_mf}. They imply that i) neither $\bf{k_1}$ nor $\bf{k_2}$ wave vectors form a multi-$\bf{k}$ structure, but rather separate phases; ii) $\bf{k_1}$ reflections are more sensitive to the applied fields then the $\bf{k_2}$ ones, this might be caused either by different anisotropy of the two magnetic structures or by peculiarities of the magneto-elastic coupling of structural twins and magnetic domains.\\
\begin{table}
\caption{Intensities of selected magnetic peaks measured in three states distinguished by cooling (CF) and measuring (MF) fields: 1) cooled in zero field and measured in zero field (CF/MF=0/0), 2) cooled in 3 T and measured in 3 T (CF/MF=3/3), 3) cooled in 3 T and measured in zero field (CF/MF=3/0). Left part corresponds to ${\bf{H}}_{1-10}$, right part to ${\bf{H}}_{1-11}$.}
\label{tab_dom_mf}
\centering
\small
\begin{tabular}{|c|ccc|c|ccc|}
\hline
\diagbox{hkl}{CF/MF}& 0/0&  3/3&  3/0&\diagbox{hkl}{CF/MF}& 0/0& 3/3& 3/0\\
\hline
{\threehalf} {\threehalf} 0&62(2)&25(5)&36(2)&{\threehalf} {\threehalf} 0&43(2)&0&17(1)\\
0 {\threehalf} {\threehalf}&54(2)&17(1)&31(2)&0 {\threehalf} $\pm${\threehalf} &47(2)&40(2)&41(2)\\
{\threehalf} 0 -{\threehalf}&60(2)&23(1)&31(2)&$\pm$({\threehalf} 0 $\pm${\threehalf}) &47(2)&47(2)&48(2)\\
 \hline
{\threehalf} 1 0&85(3)&87(3)&86(3)&{\threehalf} 1 0&58(2)&39(2)&40(2)\\
{\threehalf} 0 -1&91(3)&87(3)&86(3)&$\pm$({\threehalf} 0 $\pm$1) &56(2)&56(2)&58(2)\\
0 {\threehalf} 1&88(3)&77(3)&76(3)&1 {\threehalf} 0 &60(2)&39(2)&45(2)\\
 & & & &$\pm$(1 0 $\pm$ {\threehalf}) &62(2)&63(2)&64(2)\\
\hline
\end{tabular}
\end{table}
As  presented in Section \ref{Sec2Sub5}, even extended sets of integrated intensities were not sufficient to determine uniquely the magnetic structure - several different structures gave the same intensity distribution. Thus we performed spherical neutron polarimetry (SNP) experiments with the cryopad device on IN12 at ILL ($\lambda$=3.70 \AA) and
the mupad device on TASP at SINQ ($\lambda$=3.14 \AA).
This method allows the separation of nuclear, magnetic and magnetic chiral contributions and is very sensitive to the direction of magnetic moments (see Appendix for the details).
We measured several reflections for several crystal orientations: $\bf{k_1}$ reflections - in the ($hk0$) and ($hhl$) horizontal scattering planes, $\bf{k_2}$ reflections  - in the ($hk0$) plane.\\
\subsection{Magnetic structure determination}
\begin{figure}
\centering 
\includegraphics[width=0.95\columnwidth]{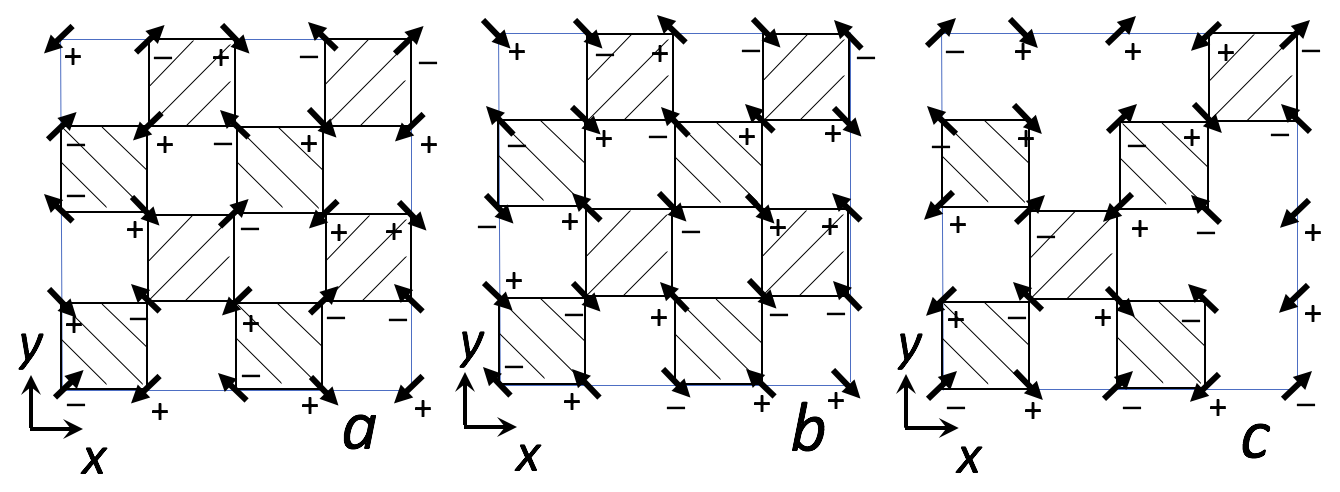} 
\caption{Magnetic models for the ($\bf{k_1}$= ({\half} {\half} 0) (a, b) and $\bf{k_2}$= (1 0 {\half}) (c) wave vectors. Tetrahedra are shown as dashed squares. The arrows present the direction of the [110] and [1-10] components of the magnetic moment, while the sign near the arrow corresponds to the direction of the $c$-component.}
\label{fig_k} 
\end{figure}

{\label{Sec2Sub5}}
In order to solve the {\MCO} magnetic structure, representation analysis was performed for
the $I4_1/amd$ and $Fddd$ space groups. However, no satisfactory fits could be obtained with the corresponding
irreducible representations (IRRs). The spin structures we propose below cannot be described by multiple IRRs either, as they break the $I$- and $F$- lattice translations. We employed the bottom-up approach similar to that for {\ZCO}\cite{Lee2008}: all possible spin arrangements were firstly constructed and then compared with the diffraction pattern. In this approach, several constraints were imposed for the magnetic structure: one is that each tetrahedron has zero total moment, which is compatible with the perturbative role of the spin-lattice coupling; the second constraint is that spins point along the [110] and [1-10] diagonals in the ${ab}$-plane, similar to that of {\ZCO} \cite{Lee2008}; and finally an equal moment value for each Cr$^{3+}$ ion was assumed.\\
As the first step, we built all 768 possible $\bf{k_1}$=({\half}  {\half} 0) long-range ordered states satisfying these constraints.
After removing the $Fd3m$ symmetric duplicates, two models - one collinear and one coplanar with 90-deg aligned spins - were found to fit the DMC powder diffraction data. They also provided a good fit to the integrated intensities collected from single crystals. Under the assumption of equal distribution of the $\bf{k_1}$ domains and equal amount of $\bf{k_1}$ and $\bf{k_2}$ phases
the refined moment size is only 1.94(3) $\mu_B$, while 3$\mu_B$ is expected from the Cr$^{3+}$ spin.\\
The same approach was employed for the $\bf{k_2}$ structure determination. 1112 structures fulfilling the above listed constraints were constructed. Three structures - one collinear and two coplanar with 90-deg aligned spins - were compatible with powder data and single crystal integrated intensities. The refined moment value is 2.08(3)$\mu_B$.\\
Despite the satisfactory fit for the diffraction data, we found that none of the obtained structures is compatible with our spherical neutron polarimetry (SNP) measurements. This can be most directly seen from the $P_{yy}$ element for the $\bf{k_2}$ reflections listed in Table \ref{tab_SNP_k2} of the Sec. \ref{Append}. For the reflections (1 k 0) with k = {\half}, {\threehalf}, and {\fivehalf} the $P_{yy}$ element for {\ZCO} monotonously increases form $-0.43$ to $-0.09$, which is consistent with the assumption that spins are in the $ab$-plane.\cite{Lee2008} While for {\MCO}, $P_{yy}$ first increases from -0.30 to -0.02, and then decreases to $-0.16$. Such a non-monotonous evolution of the $P_{yy}$ element indicates that, in contrast to {\ZCO}, spins in {\MCO} should have finite out-of-plane components.\\ 
To fit the SNP matrices, we first applied a uniform rotation of the spins. The Euler angles  $\phi, \theta, \psi$, which represent the successive rotation around the $z$, $x$ and $z$ axes, in the ZXZ convention were used as fitting parameters. The summation of the absolute difference between the measured and calculated $P_{yy}$ elements of the six measured $\bf{k_2}$ reflections was used as the goodness-of-fit criterion.
Only the structure with the [110] and [1-10] components canted 90 deg shown in Fig. \ref{fig_k}c is in agreement with the measured matrices.
Satisfactory fits were achieved with $M_c$/$M_b$=-0.378(5) and two orientation domains having the  opposite $M_b$ components, i.e.  $(M_a, M_b, M_c) \rightarrow (M_a, -M_b, M_c)$. The corresponding SNP matrices are listed in Table \ref{tab_SNP_k2} in the Sec. \ref{Append}. The fit of this model to the integrated intensities dataset is also good with the agreement factor $R_f$=5.6.\\
For the $\bf{k_1}$ structure even the SNP data are not sufficient to distinguish the two models presented in Fig. \ref{fig_k}a, b. The measured and calculated SNP matrices are listed in Tables \ref{tab_SNP1_k1}, \ref{tab_SNP2_k1} in the Sec. \ref{Append}. The refinement of the integrated intensities is equally good, the agreement factor is $R_f$=11.3.
The model with the 90-deg arranged [110] and [1-10] components has $M_c$/$M_b$= -0.376(5) and is shown in Fig. \ref{fig_k}a. To fit the SNP data we had to add contributions of two configuration domains related by the transformation $(M_a, M_b, M_c) \rightarrow (M_b, M_a, M_c)$.
The collinear structure presented  in Fig. \ref{fig_k}b gives an equally good fit to all experimental data.
For SNP the orientation domains related by the transformation $(M_a, M_b, M_c) \rightarrow (M_a, M_b, -M_c)$ should be considered. For a good integrated intensity fit ($R_f$=11.3)
we allowed the overlap of ($hkl$) and ($k-hl$) reflections of the same structure (i.e. twinning by the 4-fold axis). For one domain of the uniaxial magnetic arrangement the neutron intensity distribution in the momentum space is very anisotropic - reflections orthogonal to the easy axis are strong and reflections along the easy axis have no intensity. Summing of the contributions of the 90-deg rotated twin species leads to a more uniform intensity distribution, similar to the one from the 90-deg arrangement of the  [110] and [1-10] components.
%
\subsection{Magnetic excitations}
\begin{figure}
\includegraphics[width=0.960\columnwidth,keepaspectratio=true]{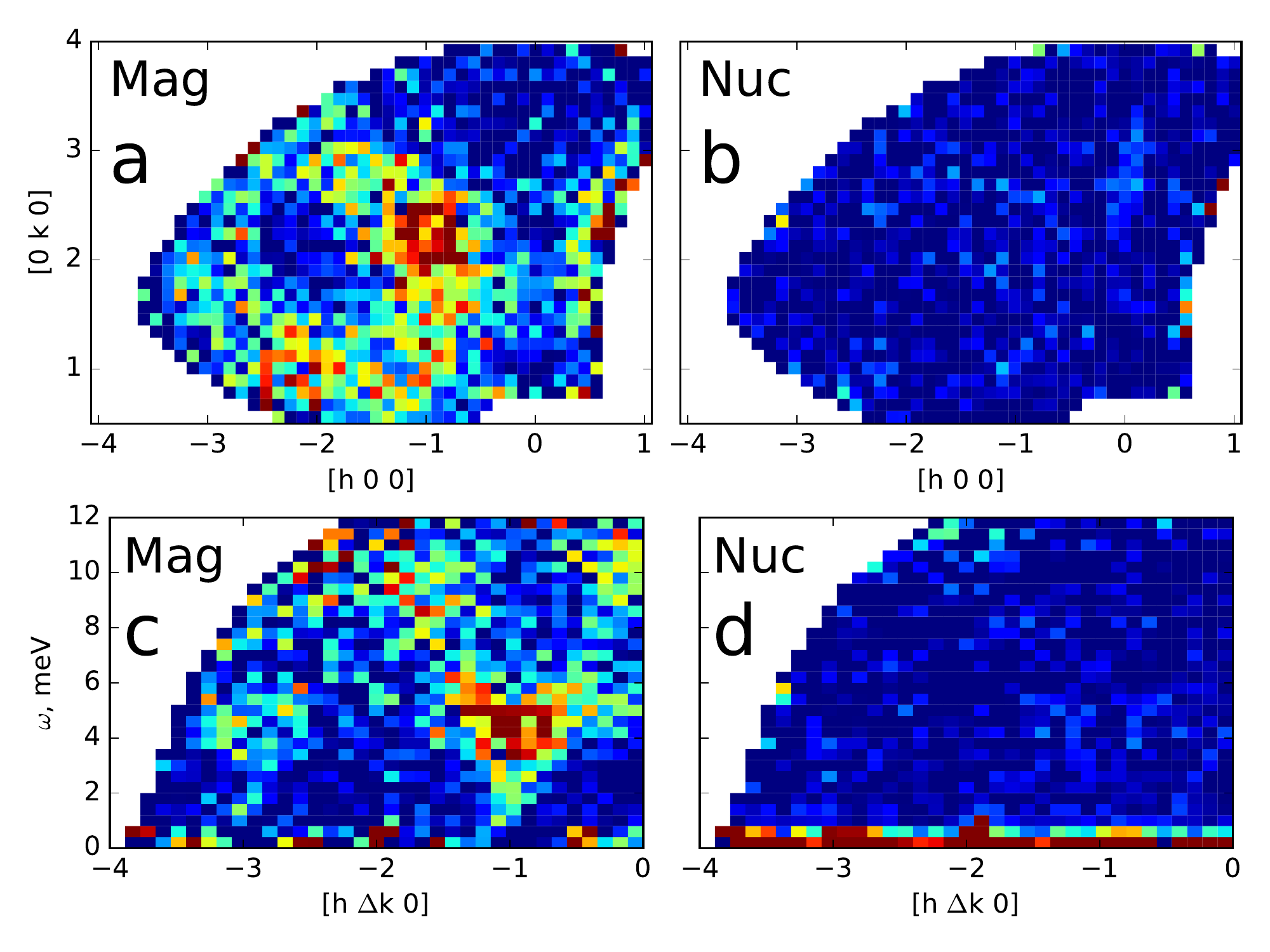}
\caption {Polarized neutron HYSPEC data at 1.5 K. Top: equal-energy slices of magnetic (a) and nuclear (b) cross sections at $\omega$= 4.5 meV (3.5 meV$<\omega<$5.5 meV). Bottom: magnetic (c) and nuclear (d) excitation spectrum along $hk0$ with $k$ summed over 1.75$<\Delta$k$<$2.25. Magnetic cross sections are obtained by summing spin-flip and non-spin-flip magnetic channels.}
\label{fig_HyspecP}
\end{figure}

\begin{figure}
\includegraphics[width=0.49\columnwidth,keepaspectratio=true]{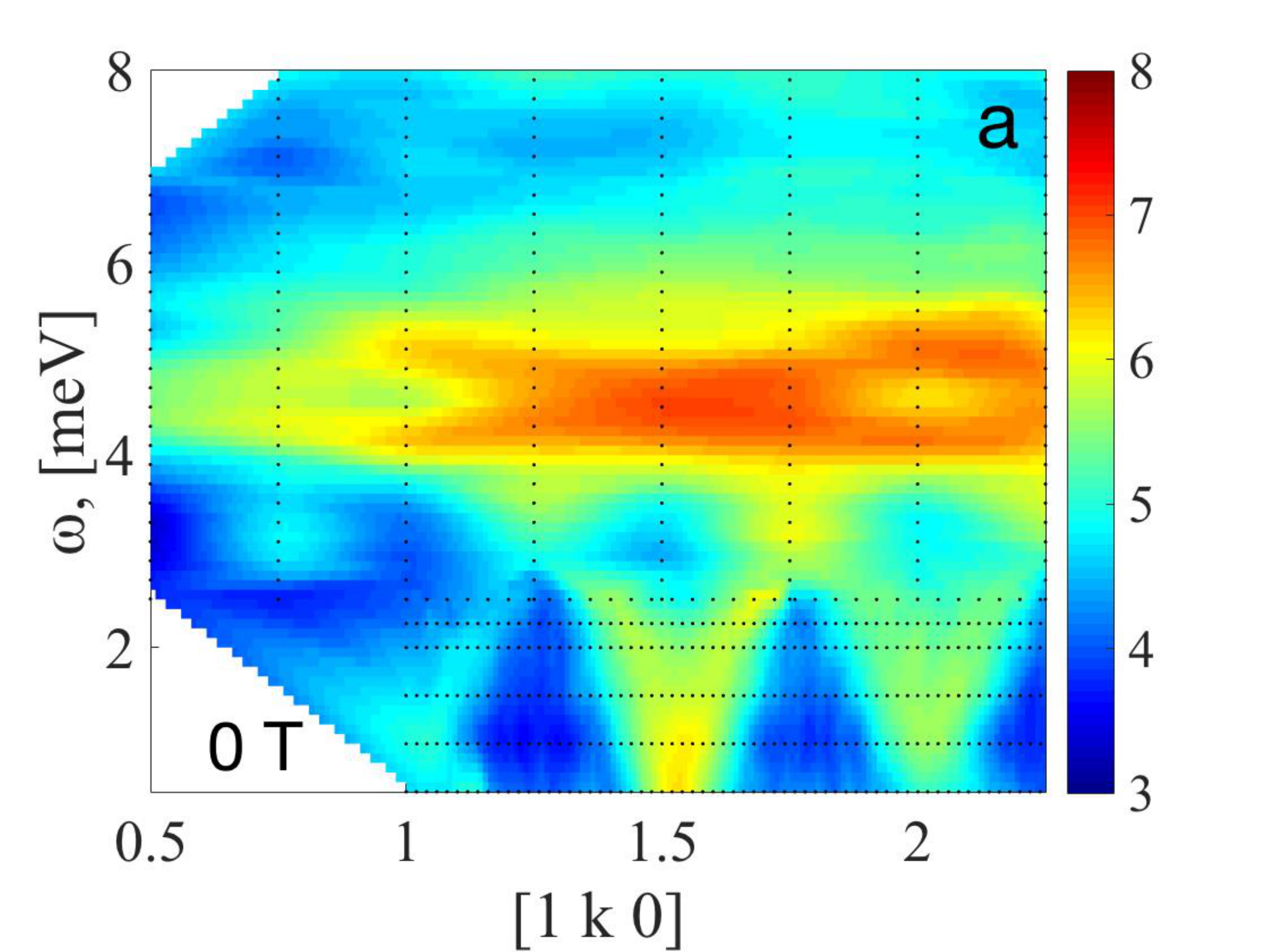}
\includegraphics[width=0.49\columnwidth,keepaspectratio=true]{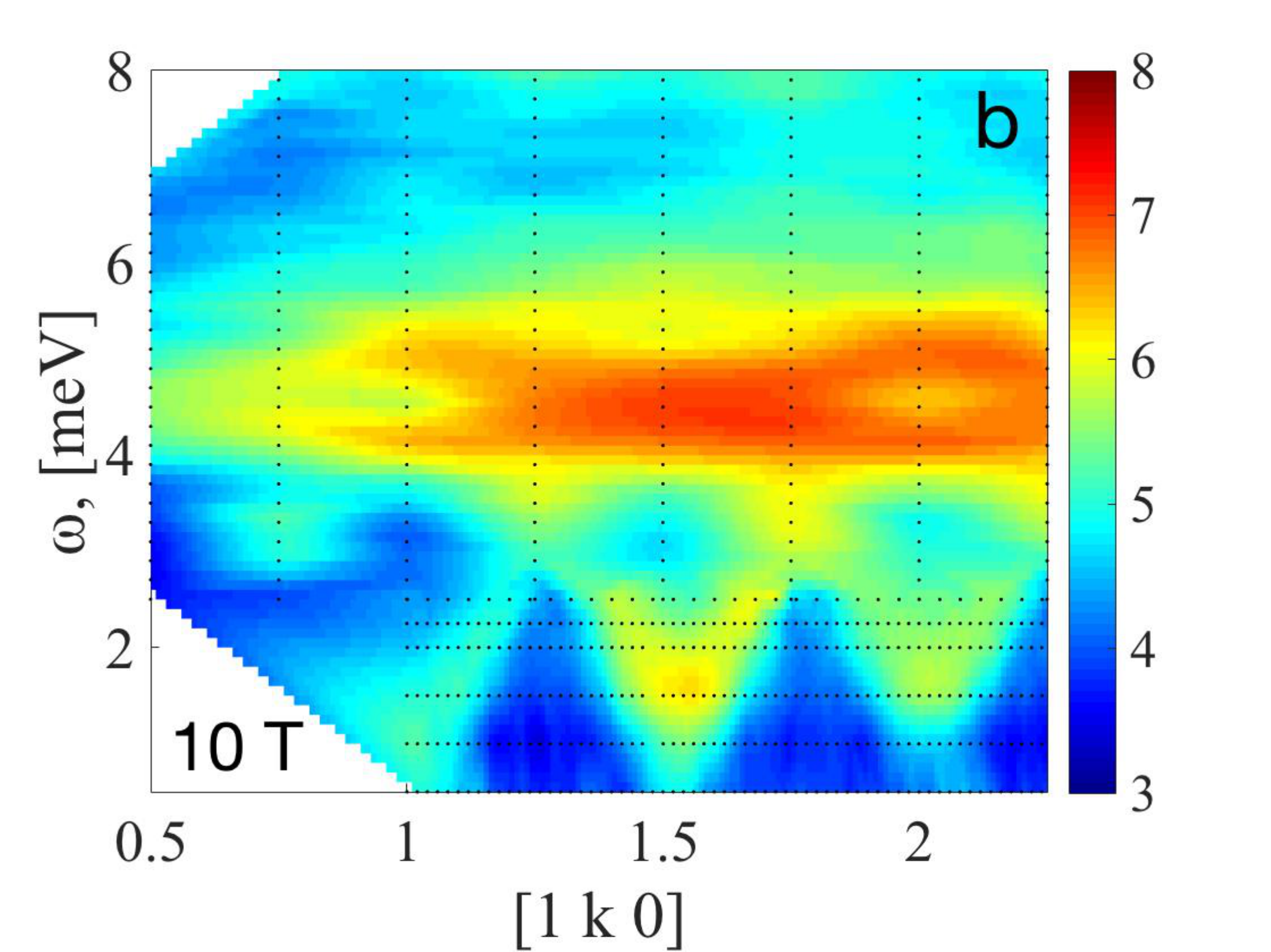}
\caption {S($Q,\omega$) maps measured on IN12 at 1.5 K: a) H= 0 T, b) ${\bf{H}}_{001}$= 10 T.}
\label{fig_IN12}
\end{figure}

{\label{Sec2Sub6}}
We measured the excitation spectrum of a 250 mg crystal-array of {\MCO} aligned with the [-211] axis vertical, using the TASP and EIGER neutron triple-axis spectrometers at SINQ; and of the 2 g crystal-array aligned with the [001] axis vertical on the triple-axis spectrometer IN12 at ILL  and the time-of-flight HYSPEC spectrometer at SNS. On the triple-axis spectrometers we used the conventional setups with PG monochromators and analysers configured with fixed final momentum $k_f$ =2.66 \AA$^{-1}$ on EIGER, 1.4 \AA$^{-1}$ and 1.17 \AA$^{-1}$ on TASP, 1.6 \AA$^{-1}$ on IN12. For the HYSPEC the initial momentum $k_i$ was fixed to 3.1 \AA$^{-1}$ and the Fermi Chopper run at the frequency $f$=240 Hz. 
Additionally we used the polarized setup on HYSPEC \cite{Winn2015} with a Heusler monochromator and a supermirror analyzer operating with $k_i$ = 2.69 \AA$^{-1}$, $f$=180 Hz. A flipping ratio of 14 was reached.
On IN12 the sample was mounted in a vertical magnet and INS data were collected in 0 T and 10 T.\\
\begin{figure}
\includegraphics[width=0.96\columnwidth,keepaspectratio=true]{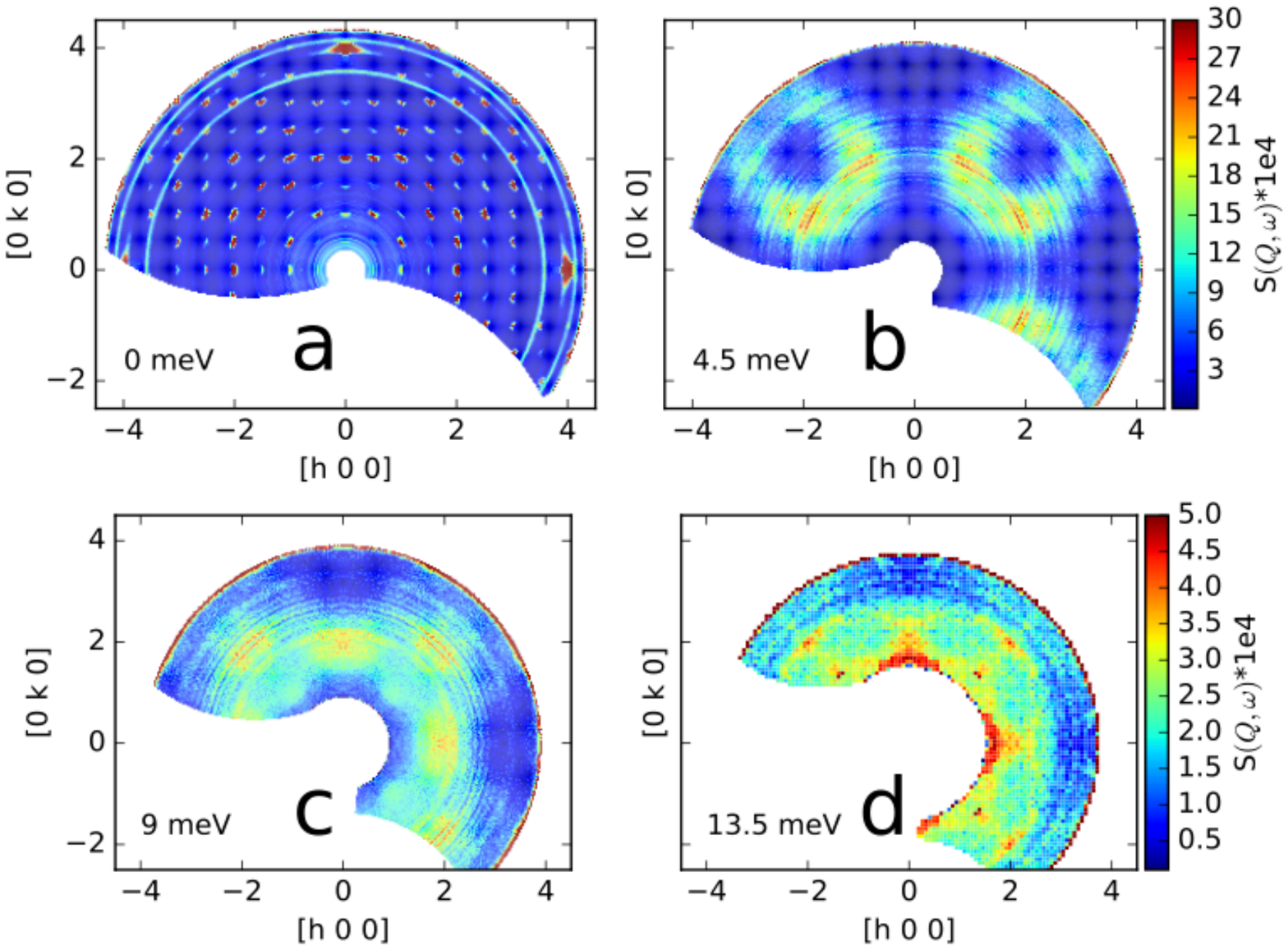}
\caption {Constant-energy slices of the S($Q,\omega$) HYSPEC data. a) $\omega$= 0 meV summed within -1 meV$<\omega <$1 meV, b) $\omega$= 4.5 meV (3.5 meV$<\omega<$5.5 meV), c) $\omega$= 9 meV (7.5 meV$<\omega<$11 meV), d) $\omega$= 13.5 meV, (12.5 meV$<\omega<$14.5 meV).}
\label{fig_HyspecQ}
\end{figure}
\begin{figure}
\includegraphics[width=0.96\columnwidth,keepaspectratio=true]{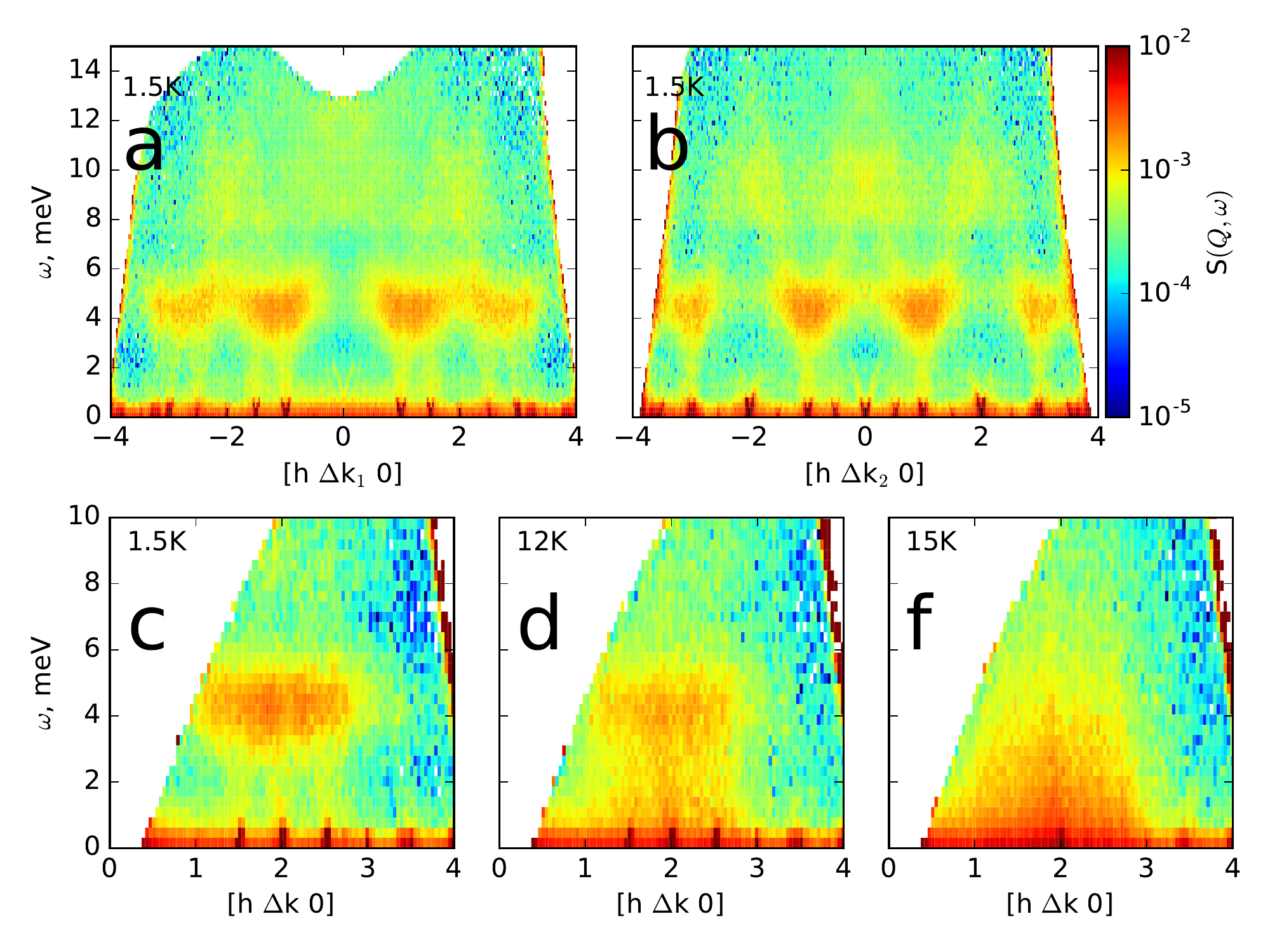}
\caption {Excitation spectrum S($Q,\omega$) from HYSPEC data. Top: 1.5 K slices along $[$h k 0$]$ with k summed over $\Delta$k$_1$=[1.3, 1.7] (a), $\Delta$k$_2$=[1.8, 2.2] (b). Bottom: temperature dependence of the S($Q,\omega$) slice along $[$h k 0$]$ with $\Delta$k=[0.8, 1.2] at 1.5 K (c), 12 K (d), 15 K (f).}
\label{fig_HyspecE}
\end{figure}
The measured spectra are very similar for the two samples but for the large mass sample the INS signal is 
significantly higher, so in Figs. \ref{fig_HyspecQ}-\ref{fig_IN12} we present the HYSPEC and IN12 data.
The INS spectra are dominated by the resonance modes centered at 4.5 and 9 meV. Within the resonances the intensity is strongly modulated with the momentum. The measured constant-energy slices (Fig. \ref{fig_HyspecQ}) are similar to the ones published in Refs. [\onlinecite{Lee2002}, \onlinecite{Tomiyasu2013}] and resemble spin correlations within a hexagon and a heptamer. The energy-momentum cuts (Fig. \ref{fig_HyspecE}) are novel, they highlight a new aspect - the resonances have well defined dispersive $Q,\omega$-boundaries and extend over ca. 2 meV. They are intrinsically broader than the resolution of the HYSPEC setup ($\delta E_{0meV}$=0.88 meV, $\delta E_{5meV}$=0.61 meV, $\delta E_{9meV}$=0.42 meV).
Steep dispersive spin waves start at magnetic peaks of the multiple wave vectors with small excitation gaps. The gaps are very similar for different wave vectors, i.e. $\Delta_{\bf{k_1}}$=0.80(4) meV and  $\Delta_{\bf{k_2}}$=0.67(4) meV (from TASP measurement), and are compatible with the easy-plane anisotropy revealed by ESR. \cite{Yoshida2006, Glazkov2009} The dispersive modes smoothly enter the resonance bands, which in turn are also weakly dispersing. This dispersion is very clearly visible in the S($Q,\omega$) cuts measured on IN12 (Fig. \ref{fig_IN12}). Neither feature changes significantly in the ordered state, but then both soften simultaneously and abruptly, close to $T_N$=12.5 K. Clearly they are intrinsically connected.\\
We verified the magnetic origin of both the excitation features by using XYZ-polarization analysis\cite{Moon1969} on HYSPEC. Under the assumption of isotropic magnetic scattering the magnetic ${\DSC}_{mag}$, incoherent ${\DSC}_{inc}$ and nuclear ${\DSC}_{nuc}$ cross sections were evaluated by the following equations\cite{Stewart2008}:
\begin{eqnarray}
{\DSC}_{mag} &=& 2{\DSC}^x_{sf}+2{\DSC}^y_{sf}-4{\DSC}^z_{sf}
\nonumber\\
{\DSC}_{mag} &=& 4{\DSC}^z_{nsf} -2{\DSC}^x_{nsf}-2{\DSC}^y_{nsf}
\nonumber\\
{\DSC}_{inc}&=& {\threehalf}(3{\DSC}^z_{sf}-{\DSC}^x_{sf}-{\DSC}^y_{sf})
\nonumber\\
{\DSC}_{nuc}&=& {\DSC}^z_{nsf} -{\half}{\DSC}_{mag} -{\third}{\DSC}_{inc}
\nonumber
\label{crossec}
\end{eqnarray}
where $x, y, z$ refer to the direction of the incident polarization, $sf$ and $nsf$ stands for spin-flip and non-spin-flip. The magnetic cross section 
presented in Fig. \ref{fig_HyspecP}c contains both spin waves and resonances, while the nuclear (Fig. \ref{fig_HyspecP}d) and incoherent cross sections contain only background. We could not identify any contribution of phonon or hybridized spin-phonon excitations to the resonance modes.\\
Lastly, we tested the response of the excitations to a magnetic field along [001] (Fig. \ref{fig_IN12}).The gaps of dispersive spin waves increase from $\approx$0.75 meV at 0 T to $\approx$1.5 meV at 10 T, such behaviour is expected for a conventional AF. However, the 4.5 meV resonance shows no significant changes.\\
\section{Theoretical description}{\label{Sec3}}
\subsection{The status of the effective Hamiltonian approach}{\label{Sec3Sub1}}
The current status of the theoretical comprehension of the {\ACO} chromites can be shortly summarised as follows.
When the resonance modes in the {\ACO} chromites were discovered, the Heisenberg antiferromagnet on the pyrochlore lattice (HAFP) model was anticipated to describe these excitations. 
The HAFP model has a non long-range ordered ground state, which is however strongly correlated. Such ground state is highly nontrivial and is rooted in connectivity and frustration of the pyrochlore lattice. The system fluctuates between configurations with zero net magnetic moment on each tetrahedron.
These low-frequency fluctuations (zero-energy modes) cost no energy and enable the system to wander from one GS to another without leaving the manifold. 
 Yet, the fluctuations are not completely random as tetrahedra share corners, and these correlations give rise to sharp features, termed pinch points, in a diffraction pattern.\cite{Moessner1998}\\
The pinch points were not observed in {\ACO}, so additional terms, such as a further neighbor exchange or spin-lattice (SL) coupling, were examined.\cite{Conlon2010, Tchernyshyov2002, Penc2004, Bergman2006}
Several approaches implementing the SL coupling should be mentioned here. The SL coupling was mapped as a quadric term in the free energy expansion \cite{Tchernyshyov2002}, as an effective biquadratic interaction term \cite{Penc2004} or implemented in the site-phonon model.\cite{Bergman2006}
These models successfully elucidated the plateau at the half of the saturation magnetization of the {\ACO} chromites, but depending on the model and its parameter choice, different ground states and emerging excitations were found. The complicated details of the experimentally determined long-range ground states and admixture of long- and short- range excitations are not predicted by these models.\\
We could not explain our experimental results by starting with the HAFP Hamiltonian with further neighbor or effective SL couplings. The ground states of such Hamiltonians were inconsistent with the experiments and linear spin wave calculations based on these ground states gave multiple dispersive branches instead of the single branch observed experimentally, and equally spaced resonance modes were not obtained. New theoretical approaches to the problem will be very useful.
\subsection{Cluster calculations}{\label{Sec3Sub2}}
There is abundant experimental evidence that the resonance modes are a persistent feature of spin correlations in the {\ACO} chromites independent of the fine details of the ground state. 
We extended the idea of Tomiyasu $et~al.$\cite{Tomiyasu2013} about the cluster modes with classical spins and developed a quantum spin model of two corner-sharing tetrahedra (TCST).
We used an analytical approach\cite{Haraldsen2011, Haraldsen2016} simplifying the isotropic Heisenberg Hamiltonian for large spin clusters and facilitating derivation of its energy eigenstates and eigenvalues.
For a large cluster of spins the Hamiltonian cannot be solved analytically due to a large dimension of the Hilbert space. The number of states for a system with $n$-particles of spin $S$ increases as (2$S$+1)$^n$. 
In such cases the cluster can be decomposed into subgeometries, which maintain the exchange symmetry of the initial cluster. The excitations of the large cluster 
are described through the excitations of the subgeometries, where the functional form of a cluster structure factor is not dependent on the spin value,\cite{Haraldsen2015} but on the 
individual subgeometries.\cite{Haraldsen2011} An important consequence of this is that transitions between discrete energy levels of the large cluster can be calculated using eigenfunctions of 
the subgeometries. This allows direct comparison between our calculations and the measured INS spectra.\\
\begin{figure}
\includegraphics[width=0.40\columnwidth,keepaspectratio=true]{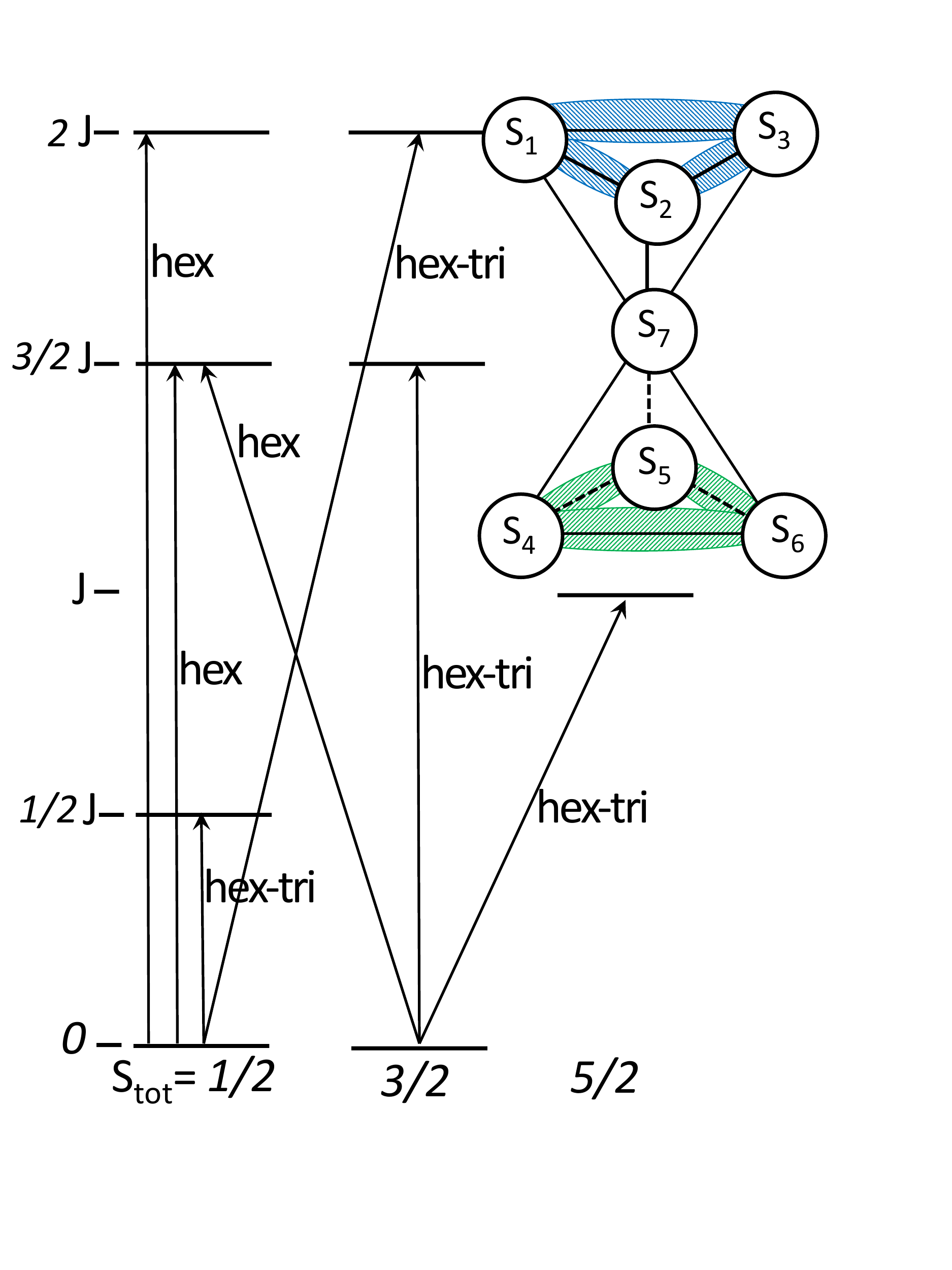}
\caption {$S$ = {\threehalf} TCST cluster and its levels. The states are indicated in the heptamer $S_{i=1-7}$, hexamer $S_{i=1-6}$ and two trimer  bases. The $S_{\triangle_1}=S_1-S_2-S_3$ trimer is shown in blue, $S_{\triangle_2}=S_4-S_5-S_6$ trimer - in green.}
\label{fighep1}
\end{figure}
We consider a $S$ = {\threehalf} TCST cluster with the antiferromagnetic exchange $J$ presented in Fig.~\ref{fighep1}. In the individual spin representation this cluster has ($2\cdot{\threehalf} +1)^7$=16384 total states. The number of states is significantly reduced when choosing the following subgeometries - two trimers: $S_1-S_2-S_3$ (denoted as $S_{\triangle_1}$) and $S_4-S_5-S_6$ ($S_{\triangle_2}$), and a hexamer $S_{hex}$ containing spins $S_{i=1-6}$. 
The Heisenberg Hamiltonian written through the basis sets of these subgeometries is
\begin{equation}
H=J\{(S_1 \cdotp S_2 + S_1 \cdotp S_3 + S_2 \cdotp S_3)
+ (S_4 \cdotp S_5 + S_4 \cdotp S_6 + S_5 \cdotp S_6) +
 \sum^6_i S_i \cdotp S_7 \}.
\end{equation}
The energy eigenstates of the TCST cluster are then
\begin{equation}
E=\frac{J}{2}[S_{tot}(S_{tot}+1)-S_{hex}(S_{hex}+1)+S_{\triangle_1}(S_{\triangle_1}+1)+S_{\triangle_2}(S_{\triangle_2}+1)- \sum^7 S_j (S_j +1) ],
\end{equation}
where $S_{tot}$ is the total spin state of the system and $S_j$ are the individual $S$= {\threehalf} spins.\\
\begin{figure}
\includegraphics[width=0.70\columnwidth,keepaspectratio=true]{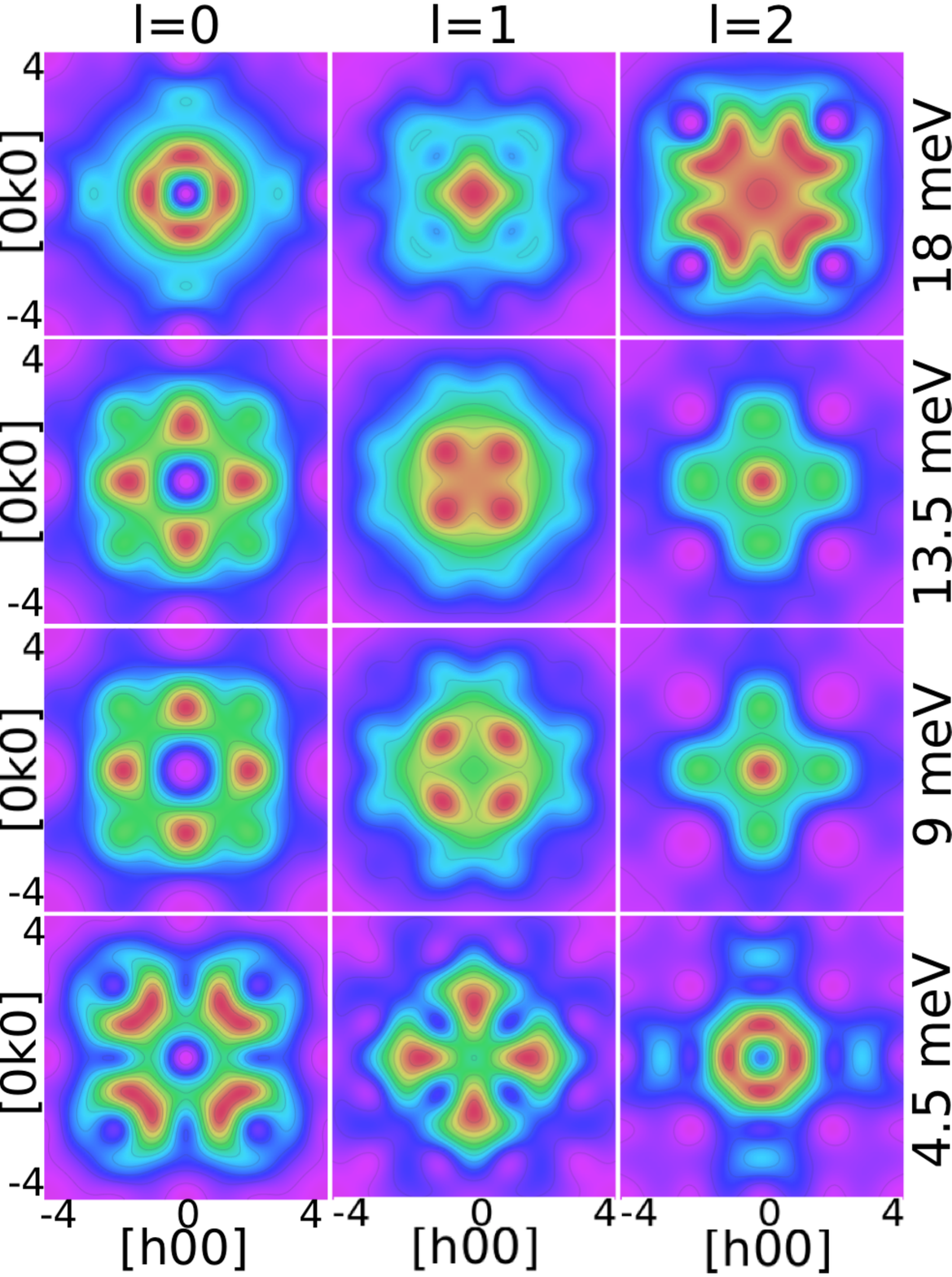}
\caption {Squared and orientation averaged structure factors for the first four excitations of the TCST cluster. The S($Q,\omega$) maps are organised in a ($i, j$)-grid with the $i$-column index increasing with the energy of the excitation and the $j$-raw index increasing with the $l$-index of the $hkl$ plane.}
\label{fighep2}
\end{figure}
Fig. \ref{fighep1} shows the lowest energy cluster levels, as well as their $S_{tot}$ designations and eigenstates. 
The ground state of the cluster is a doublet consisting of $S_{tot}$={\half} and $S_{tot}$={\threehalf} states. Other energy levels are equally spaced with the $J$/2 interval.\\
To calculate the $Q$-dependence of the INS structure factor we reduced the $S$={\threehalf} cluster to the $S$={\half} analogue\cite{Haraldsen2015} and focused on
an 'exclusive structure factor'\cite{Haraldsen2016} for the excitations within a specific magnetic multiplet of final states $\mid\Psi_f(\lambda_f)\rangle$ from the given 
initial state $\mid\Psi_i\rangle$:
\begin{equation}
S^{(fi)}_{ba}(\vec{q})= \sum_{\lambda_f}\langle \Psi_i\mid V^\dagger_b \mid\Psi_f(\lambda_f)\rangle
\langle \Psi_f(\lambda_f)\mid V^\dagger_a \mid\Psi_i\rangle
\end{equation}
where the vector $V(\vec{q})$  is a sum of spin operators over the cluster:
\begin{equation}
V = \sum_{\vec{x}_i} S(\vec{x}_i) e^{i\vec{q}\cdotp\vec{x}_i}.
\end{equation}
To obtain the functional form the spatial indices $a, b$ could be reduced to $z$.\\
Using this procedure we determined the INS structure factors for the four lowest-energy transitions.
The 4.5 meV excitation ($\Delta$E = $J$/2) is a  $S_{tot}$ = {\half} to  $S_{tot}$= {\half} transition, it can be presented  through the heptamer, hexamer and trimers basis  ($\mid S_{tot}S_{hex}S_{\triangle_1}S_{\triangle_2}\rangle$) as
($\langle {\frac{1}{2}} 1 {\frac{1}{2}}\mid {\frac{1}{2}}2{\frac{1}{2}}{\frac{3}{2}}\rangle$). 
The 9 meV excitation ($\Delta$E = $J$) is a $S_{tot}$ = {\threehalf}
to $S_{tot}$= {\fivehalf} transition (($\langle {\frac{5}{2}} 4 {\frac{3}{2}} {\frac{5}{2}}\mid {\frac{3}{2}} 3{\frac{3}{2}}{\frac{3}{2}}\rangle$)), 
which also involves the hexamer and trimer bases. The third and fourth excitations ($\Delta$E = {\threehalf}$J$ and $\Delta$E = 2$J$, respectively)
consist of a combination of multiple excitations that encompass excitations of hexamers and heptamers.
\\
Finally, to compare the $Q$-dependence of the TCST cluster with the {\MCO} INS spectra, we summed 
the squared structure factors for all possible orientations of the cluster on the pyrochlore lattice.
The match between the calculated (Fig. \ref{fighep2} $i$=1,$j$=1,3) and measured (Fig. \ref{fig_HyspecQ}  b-d) intensity distributions of the three lowest resonances at 4.5 meV, 9 meV and 13.5 meV is remarkable.
We therefore think that the resonances are rooted in the quantum levels of the cluster.\\
Furthermore, due to the averaging of the multiple cluster configurations through the pyrochlore lattice, we
expect that effects of magnetic field applied in one crystal direction will be diminished, which is consistent with our experimental observations.
\section{Summary and Discussion}{\label{Sec4}}
In order to understand the origin of resonances in the {\ACO} spinels we performed a detailed experimental study of {\MCO}. We confirm the simultaneous magnetic and structural transition at $T_N$=12.5 K. We observe splitting of the cubic reflections which can be explained by tetragonal symmetry, but the fine details of the low-temperature crystal structure observed in several powder and single crystal synchrotron x-ray diffraction experiments are not consistent. In some experiments we observe weak reflections with  ${\bf{k}}$=({\half} {\half} {\half}) and rather strong intensities at the $P$-lattice positions, but in other experiments they are absent. Presently we give preference to the tetragonal model of Ehrenberg $et~al.,$\cite{Ehrenberg2002} as it explains the details that are consistent in all our x-ray diffraction experiments. It is important to perform a state-of-art diffraction experiment meeting the challenge of simultaneous measurement of a sufficient set of weak superstructure reflections ($\approx 10^{-3}$ weaker than the main peaks) and resolving the splitting of the cubic reflections ($\Delta a/a\approx 10^{-3}$). It will be important to study the microstructure in the LT phase and its consequence on the magnetic orders by means of electron transmission microscopy. It is well documented\cite{Carter1987a, Carter1987b} that spinels, besides the inversion, have tendency for complicated grain and twin boundaries with dislocations and cation rearrangement and the variations of the HT/LT structural transition might germinate from these effects.\\
By neutron diffraction we confirm two main magnetic propagation vectors $\bf{k_1}$=({\half} {\half} 0) and $\bf{k_2}$=(1 0 {\half}) below $T_N$. 
The $\bf{k_1}$ and  $\bf{k_2}$ reflections respond differently to magnetic fields, this implies formation of multiple domains and not of multi-k structures. 
In zero magnetic field the ordered moment reaches only 1.94(3) $\mu_B$ for $\bf{k_1}$ and 2.08(3) $\mu_B$ for $\bf{k_2}$, when equal amount of the $\bf{k_1}$ and  $\bf{k_2}$ phases is assumed; thus the magnetic long-range order is partial. 
For the $\bf{k_1}$=({\half} {\half} 0) and $\bf{k_2}$=(1 0 {\half}) configuration arms the magnetic moments are predominantly in the ${ab}$-plane, but the finite $M_c$-components with the ratio $M_c$/$M_b \approx$ -0.375 also exist. 
The $\bf{k_2}$ structure is determined unambiguously, in the ${ab}$-plane it has the 90-deg arrangement of [110] and [1-10] components, while for the $\bf{k_1}$ structure we cannot distinguish between the diagonal collinear and 90-deg arranged [110] and [1-10] components even by combination of single crystal neutron diffraction and spherical neutron polarimetry.\\
From these multiple orderings an admixture of the dispersive spin waves and flat resonance modes emerges. As These excitations studied by inelastic neutron scattering including XYZ-polarization analysis have the magnetic origin and are inherently connected. The dispersive spin waves for the $\bf{k_1}$ and  $\bf{k_2}$ magnetic orders are quite similar. The gaps are responsive to applied magnetic fields, they increase when field is applied along  [001], while the 4.5 meV resonance mode is field-independent.\\
The resonance modes have two aspects. From one side, the resonances are weakly dispersing over 2 meV and their thermal evolution is the same as for the dispersive spin waves. Thus they behave like optical branches of spin waves. We did not detect any dynamic distortions in the form of low-energy spin-phonon contribution in the studied Q-range. They, however, could exist at higher Q and future theoretical and experimental clarifications are required.\\
We did not succeed to fit the observed excitation spectrum to the Heisenberg pyrochlore antiferromagnet model with further neighbor or effective spin-lattice couplings.
We found, however, that resonances resemble the excitations of the TCST cluster. Our cluster Hamiltonian explains the equal energy spacing and $Q$-distribution of neutron intensity of the resonances remarkably well. Obviously, the cluster model does not capture the dispersive part of the spectrum.\\
So what could be a microscopic picture of the magnetic ground state and emerging excitations?
At $T_N$ the crystal breaks down into structural twins and magnetic domains. The tiny structural distortions are long-ranged and static. They are hard to reproduce from experiment to experiment as they are controlled by microstructure of material, which changes from sample to sample.
Only part of the moment is ordered, the rest fluctuates, 
but essentially there is no zero-energy modes left in the magnetically ordered state below $T_N$. The majority of fluctuations are collective and take the form of acoustic and optical spin waves. The typical distance an optical spin wave propagates is a cluster built of two adjacent tetrahedra.
On such short distance the excitations can be better analyzed as cluster transitions: they have energy spacing of $J$/2 and their $Q$-dependence is well described by the subgeometries - trimers and hexamers. 
Longer distances of propagation of the coherent excitations lead to dispersion.
To some extent the situation is reminiscent of spinons in AF spin chains.\cite{Mourigal2013} In spin chains a local spin flip fractionalises into two domain walls, leading to a characteristic continuum in the INS spectrum. The lower and upper boundaries of this continuum are defined by the anisotropies and exchanges of the system. In the {\ACO} spinels such spin flips could be confined to TCST clusters.\\
A uniform theoretical description of the excitations in the {\ACO} spinels is still missing. The Yet, our study uncovers that the resonances are rooted in excitations of a quantum antiferromagnetic heptamer. 
A flip of classical AF hexagons depict the first excitation mode\cite{Lee2009}, classical heptamers with different  ferro- and  antiferro- couplings represent the patterns 
of the last three excitations\cite{Tomiyasu2013}; with the AF TCST unit we explain for the first time the energy and dynamic structure factors of all four resonances consistently. It should be noted that the individual excitations of heptamer are excitations of smaller cluster bases. The energy levels of the heptamer govern the system, but the structure factor provides a
fingerprint for the nature of the spin excitation depending on the excited subgeometry. 
Therefore, in the heptamer model, we can describe the cluster as the interaction between multiple subgeometries of trimers and hexamers. 
This is very important first step towards the desired complete description.\\
As the next step we envision the placement of the heptamers in a mean field (MF) of intercluster interactions and calculation of the excitation spectrum by the random phase approximation. The Zeeman term added to this mean-field Hamiltonian should explain the different behaviour of acoustic and optical spin wave branches seen experimentally. 
We anticipate that averaging of the six cluster orientations through the pyrochlore lattice will diminish the effect of uniaxial magnetic field.
The success of such MF approach is exemplified on other frustrated systems, such as the coupled
tetrahedra system Cu$_2$Te$_2$O$_5$X$_2$ (X=Cl, Br)\cite{Prsa2009} or the coupled triangle system Ba$_2$NbFe$_2$Si$_2$O$_{14}$.\cite{Stock2011, Loire2011}
The spin waves observed in these systems  could be successfully reproduced starting from the cluster units.\cite{Jensen2009, Jensen2011} 
It should be noted, however, that applicability of this approach to spinels could be challenged by a similar strength of the inter- and intra- interactions between the heptamers. An ingenious solution of this issue is required.\\ 
Possibly a strict description of such systems is beyond the conventional linear spin-wave theory and requires taking into consideration magnon decays. Such extended analysis done for the noncollinear triangular antiferromagnet lattice \cite{Chernyshev2006} yields a mixture of sharp single-magnon modes and a multi-magnon continuum. This is also one of the scenarios discussed for the strongly spin-orbital coupled system $\alpha$-RuCl$_3$.\cite{Winter2017}\\ 
Another, rather computational challenge arises from our study. For solution of the {\MCO} magnetic structure the conventional symmetry analysis breaks down and on top several different magnetic arrangements give rise to the same diffraction pattern.
This calls for development of new computational algorithms, combining information available from different techniques. Hopefully our presented work would stimulate such developments.
\section{Appendix}{\label{Append}}
The scattering of polarized neutrons is well described in Ref. [\onlinecite{Blume, Maleev, Brown01}]. The incoming $\Pin$ and scattered $\Ps$ polarization of a pure magnetic reflection is usually defined in its local coordinate system for the specific crystal orientation - $z$ is the vertical direction, $x$ is the horizontal direction of the scattering vector $\bf{q}$, $y$
completes the right-handed Cartesian set. The polarization matrix for a single domain with no chiral contribution can be written:\cite{Brown01}
\begin{equation}
P_{ij}=\left|
\begin{array}{ccc}
-1&0&0\\ 
0&\frac{-M^2+R_{yy}}{M^2}& \frac{R_{yz}}{M^2}\\
0& \frac{R_{yz}}{M^2}&\frac{-M^2+R_{zz}}{M^2} 
\end{array}
\right|
\label{eq:eq1}
\end{equation}
with $i$ - incoming,  $j$ - outcoming component of polarization, $M^2={\bf{M_{\perp}}\cdot \bf{M^*_{\perp}}}$, $R_{ij}=2\mathfrak{R}(M_{\perp i} M^*_{\perp j})$ and $M_{\perp}$ - projection of the Fourier transform of the magnetization perpendicular to ${\bf{q}}$.
\begin{figure}
\includegraphics[width=0.85\columnwidth,keepaspectratio=true]{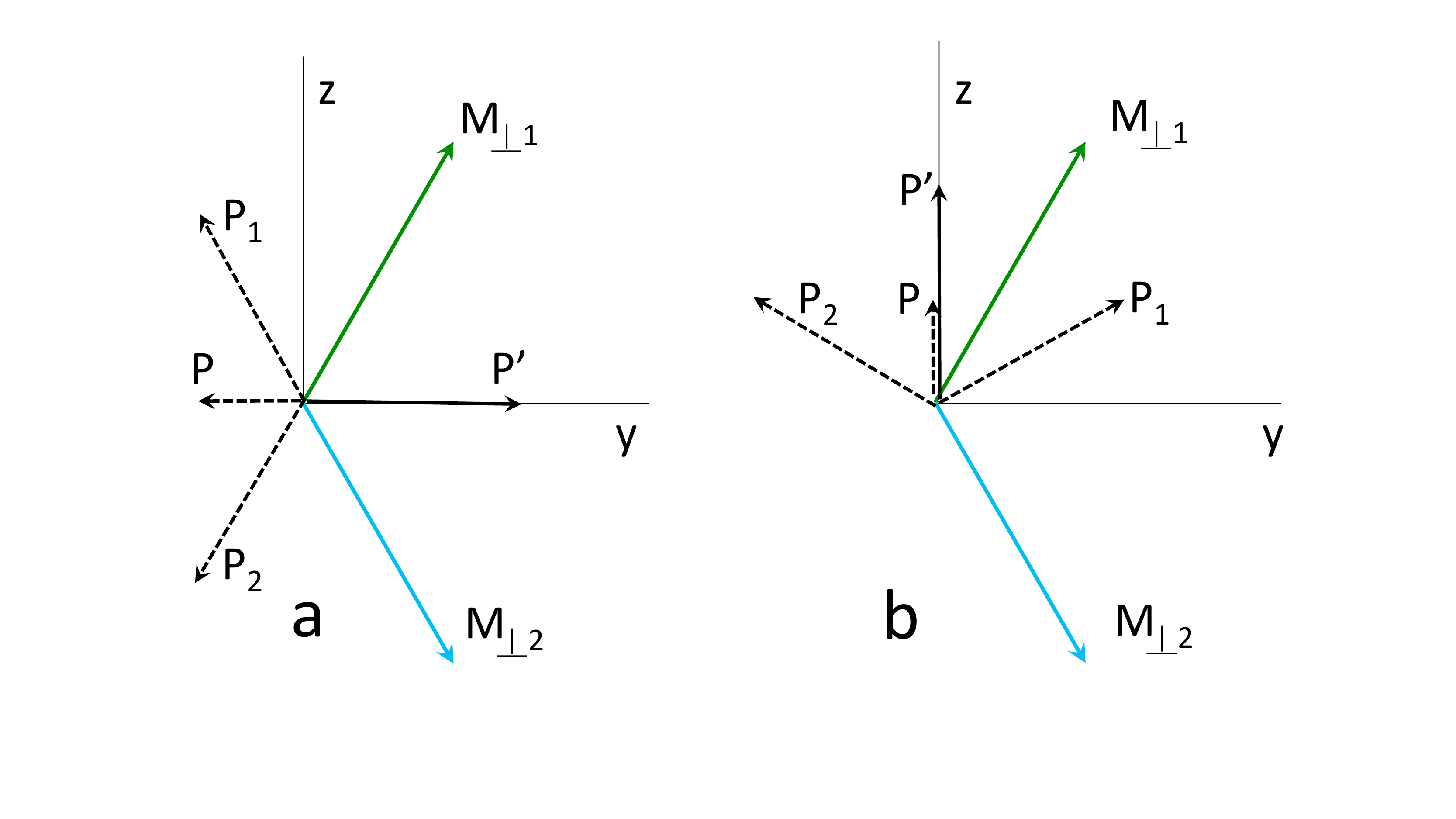}
\caption{
Depolarization of neutron beam with scattered neutron polarization $\Ps$ for incoming polarization $\Pin \parallel y$ (a) and $\Pin \parallel z$ (b) by two orientation domains with the Fourier transforms of the magnetization perpendicular to ${\bf{q}}$ denoted ${\bf{M_{\perp}}}_1$ and ${\bf{M_{\perp}}}_2$.}
\label{fig_1S}
\end{figure}
When orientation domains are present, the scattered beam is depolarized. A vectorial scatch of such depolarization is presented in figure~\ref{fig_1S}. If two orientation domains have opposite $z$-components of $\bf{M_{\perp}}$ (Fig.~\ref{fig_1S} a), the $\Pin$ is rotated to $\Ps_1$ and $\Ps_2$ by the two domains, respectively. The total $y$-component of $\Ps$ is reduced and for equal population of these domains will even vanish. The same depolarization happens for the $\Ps$ $z$-component.\\
\begin{table}
\caption{\label{tab_SNP1_k1} SNP matrices for ${\bf{k_1}}=({\half} {\half} 0)$ with the ($hk0$) horizontal scattering plane.}
 \begin{tabular}{c c c}
 \hline
Reflections & experimental & calculated\\
 \vspace{1mm}
$(0.5\ 0.5\ 0)$ & 
 $\left( \begin{array} {ccc} 
	-0.92(2) & -0.01(2) & 0.02(2) \\
    -0.03(2) & 0.84(2) &-0.02(2) \\    
     0.02(2) & 0.01(2) &-0.77(2)   
 \end{array} \right)$ &
 $\left( \begin{array} {ccc} 
	-0.92 & 0 & 0 \\
    0 & 0.80 & 0 \\    
    0 & 0 & -0.80  
 \end{array} \right)$ \\
 $(1.5\ 1.5\ 0)$ &
  $\left( \begin{array} {ccc} 
	-0.93(1) & 0.02(1) & -0.01(1) \\
     0.03(1) & 0.80(1) & -0.06(1) \\    
    -0.01(1) &  0.07(1) & -0.81(1)   
 \end{array} \right)$ &
 $\left( \begin{array} {ccc} 
	-0.92 & 0 & 0 \\
    0 & 0.80 & 0 \\    
    0 & 0 & -0.80   
 \end{array} \right)$ \\
 $(2.5\ 2.5\ 0)$ &
  $\left( \begin{array} {ccc} 
	-0.93(1) & 0.02(1) & -0.01(1) \\
     0.01(1) & 0.80(1) &  0.04(1) \\    
    -0.01(1) & 0.04(1) & -0.80(1)   
 \end{array} \right)$ &
 $\left( \begin{array} {ccc} 
	-0.92 & 0 & 0 \\
    0 & 0.80 & 0 \\    
    0 & 0 & -0.80   
 \end{array} \right)$ \\
 \hline
\end{tabular}
 \end{table}
\begin{table}
\caption{\label{tab_SNP2_k1} SNP matrices for ${\bf{k_1}}=({\half} {\half} 0)$ with the ($hhl$) horizontal scattering plane.}
\begin{tabular}{c c c}
\hline
Reflections & experimental & calculated\\
\vspace{1mm}
  $(0.5\ 0.5\ 0)$ & 
 $\left( \begin{array} {ccc} 
-0.92(3) &  0.06(3) &-0.05(3) \\
    -0.08(3) & -0.80(3) & 0.00(3) \\    
    -0.12(3) &  0.01(3) & 0.80(2)   
 \end{array} \right)$ &
 $\left( \begin{array} {ccc} 
	-0.92 & 0 & 0 \\
    0 & -0.80 & 0 \\    
    0 & 0 & 0.80     
 \end{array} \right)$ \\
 $(0.5\ 0.5\ 1)$ &
  $\left( \begin{array} {ccc} 
	-0.91(1) &  0.00(2) &-0.02(2) \\
    -0.06(2) & -0.89(1) & 0.08(2) \\    
    -0.02(2) &  0.04(2) & 0.87(1)   
 \end{array} \right)$ &
   $\left( \begin{array} {ccc} 
	-0.92 & 0 & 0 \\
    0 & -0.88 & 0 \\    
    0 & 0 & 0.88       
 \end{array} \right)$ \\
  $(0.5\ 0.5\ 2)$ &
  $\left( \begin{array} {ccc} 
	-0.93(0) & -0.04(1) & 0.01(1) \\
    -0.02(1) & -0.89(0) & 0.09(1) \\    
    -0.05(1) &  0.06(1) & 0.90(0)   
 \end{array} \right)$ &
   $\left( \begin{array} {ccc} 
	-0.92 & 0 & 0 \\
    0 & -0.91 & 0 \\    
    0 & 0 & 0.91       
 \end{array} \right)$ \\
   $(1.5\ 1.5\ 0)$ &
  $\left( \begin{array} {ccc} 
	-0.91(1) &  0.02(1) &-0.05(1) \\
    -0.04(1) & -0.80(1) & 0.01(1) \\    
    -0.06(1) &  0.02(1) & 0.80(1)   
 \end{array} \right)$ &
   $\left( \begin{array} {ccc} 
	-0.92 & 0 & 0 \\
    0 & -0.80 & 0 \\    
    0 & 0 & 0.80       
 \end{array} \right)$ \\
  $(1.5\ 1.5\ 1)$ &
  $\left( \begin{array} {ccc} 
	-0.93(1) & -0.02(2) &-0.06(2) \\
    -0.03(2) & -0.82(1) & 0.03(2) \\    
    -0.03(2) &  0.03(2) & 0.82(1)   
 \end{array} \right)$ &
   $\left( \begin{array} {ccc} 
	-0.92 & 0 & 0 \\
    0 & -0.82 & 0 \\    
    0 & 0 & 0.82       
 \end{array} \right)$ \\
   $(2.5\ 2.5\ 0)$ &
  $\left( \begin{array} {ccc} 
	-0.91(2) &  0.04(2) &-0.02(2) \\
    -0.05(2) & -0.78(2) & 0.00(2) \\    
    -0.01(2) &  0.02(2) & 0.78(1)   
 \end{array} \right)$ &
   $\left( \begin{array} {ccc} 
	-0.92 & 0 & 0 \\
    0 & -0.80 & 0 \\    
    0 & 0 & 0.80       
 \end{array} \right)$ \\
 \hline
 \end{tabular}
 \end{table}
 \begin{table}
 \centering
 \caption{\label{tab_SNP_k2} SNP matrices for the ${\bf{k_2}}$=(1 0 {\half}) reflections. The last column presents the $P_{yy}$ component measured for {\MCO} by Lee at al.\cite{Lee2008}} 
 \begin{tabular}{c c c c}
 Reflections & experimental & calculated & [\citenum{Lee2008}]\\
 \hline
 \multicolumn{4}{c}{($hk0$) as the horizontal scattering plane}\\
 \hline
 \vspace{1mm}
$(1\ 0.5\ 0)$ & 
 $\left( \begin{array} {ccc} 
-0.92(1) & 0.03(1) & 0.00(1) \\
    -0.05(2) &-0.27(2) & 0.03(1) \\    
    -0.02(1) & 0.06(1) & 0.30(1)   
 \end{array} \right)$ &
 $\left( \begin{array} {ccc} 
-0.92 & 0 & 0 \\
    0 & -0.30 & 0 \\    
    0 & 0 & 0.30
 \end{array} \right)$ &
$\left( \begin{array} {c} 
...\\
-0.43\\    
...
 \end{array} \right)$ \\
 $(1\ 1.5\ 0)$ &
  $\left( \begin{array} {ccc} 
-0.92(1) & 0.02(1) & -0.01(1) \\
     0.02(1) & 0.01(1) & -0.08(1) \\    
    -0.02(1) & -0.07(1) & -0.01(1)   
 \end{array} \right)$ &
 $\left( \begin{array} {ccc} 
-0.92 & 0 & 0 \\
    0 & -0.02 & 0 \\    
    0 & 0 & 0.02  
 \end{array} \right)$ &
 $\left( \begin{array} {c} 
...\\
-0.18 \\    
...
 \end{array} \right)$ \\
 $(1\ 2.5\ 0)$ &
  $\left( \begin{array} {ccc} 
-0.93(1) & 0.04(1) &  0.01(1) \\
    -0.02(1) &-0.16(1) & -0.07(1) \\    
     0.00(1) &-0.06(1) &  0.16(1)   
 \end{array} \right)$ &
 $\left( \begin{array} {ccc} 
-0.92 & 0 & 0 \\
    0 & -0.16 & 0 \\    
    0 & 0 & 0.16   
 \end{array} \right)$ &
 $\left( \begin{array} {c} 
...\\
-0.09 \\    
...
 \end{array} \right)$ \\
  $(3\ 1.5\ 0)$ &
  $\left( \begin{array} {ccc} 
-0.91(1) & 0.05(1) &  0.04(1) \\
    -0.06(1) &-0.80(1) & -0.02(1) \\    
     0.02(1) &-0.05(1) &  0.75(1)   
 \end{array} \right)$ &
 $\left( \begin{array} {ccc} 
-0.92 & 0 & 0 \\
    0 & -0.75 & 0 \\    
    0 & 0 & 0.75  
 \end{array} \right)$&
 $\left( \begin{array} {c} 
...\\
...\\    
...
 \end{array} \right)$ \\
 \hline
 \end{tabular}
 \end{table}
\begin{acknowledgments}
This work was performed at SINQ, Paul Scherrer Institute, Villigen, Switzerland with financial support of the Swiss National Science Foundation (SCOPES project IZ73Z0 152734/1, Grants Nos 200021-140862 and 200020-162626).
Work at the University of Warwick was funded by the EPSRC, UK through Grant EP/M028941/1.
The HYSPEC experiment used resources at the Spallation Neutron Source, a DOE Office of Science User Facility operated by the Oak Ridge National Laboratory. The IN12 experiment was supported by the Swiss State Secretariat for Education, Research and Innovation through a Collaboration Research Group grant.
\end{acknowledgments}
\end{document}